\documentclass[10pt]{article}
\usepackage[english]{babel}
\usepackage{graphics}
\usepackage{graphicx}
\usepackage{amsmath}
\usepackage{amsfonts}
\usepackage{amssymb}%
\begin{document}

\title{Origin of supermassive black holes}

\author{V. I.~Dokuchaev$^{1}$\thanks{dokuchaev@inr.npd.ac.ru}, \
Yu. N.~Eroshenko$^{1}$\thanks{erosh@inr.npd.ac.ru} \ and \ S.
G.~Rubin$^{2}$\thanks{sergeirubin@list.ru}}

\date{}
\maketitle

\begin{center}
{\sl\small $^{1}$ Institute for Nuclear Research of the Russian
Academy of Sciences, \\
60th October Anniversary prospect 7a, Moscow, 117319, Russia \\
$^{2}$ Moscow State Engineering Physics Institute, \\
31 Kashirskoe Sh., Moscow 115409, Russia}
\end{center}

\begin{abstract}
The origin of supermassive black holes in the galactic nuclei is
quite uncertain in spite of extensive set of observational data.
We review the known scenarios of galactic and cosmological
formation of supermassive black holes. The common drawback of
galactic scenarios is a lack of time and shortage of matter supply
for building the supermassive black holes in all galaxies by means
of accretion and merging. The cosmological scenarios are only
fragmentarily developed but propose and pretend to an universal
formation mechanism for supermassive black holes.
\end{abstract}

\tableofcontents

\section{Introduction}

The physics of Black Holes (BHs) is the most developed branch of
general relativity. There were elaborated numerous mechanisms for
the formation of BH of different mass-scales and proposed a list
of convincing observational signatures in favor of BH existence. A
possible range of BH masses in the Universe is thought to be
extremely wide: from the Plank mass, $M_{\rm Pl}\sim10^{-5}$~g, up
to the huge mass of the most luminous quasars, $M_{\rm BH}\sim
10^{10}{\rm M}_{\odot}$. Conventionally we will distinguish the
stellar-mass BHs with mass in the range $\sim3-100\,{\rm
M}_{\odot}$, the Intermediate Mass Black Holes (IMBHs) with mass
in the range $\sim10^{3}-10^{5}{\rm M}_{\odot}$ and the
Supermassive Black Holes (SBHs) with mass $\geq10^{6}{\rm
M}_{\odot}$. The stellar-mass BHs are rather definitely originated
from the collapse of exhausted massive stars during supernova
explosions. At the same time the origin of other types of BHs
nowadays is quite uncertain.

Modern astrophysics provides strong indications of the existence
of BHs as among the remnants of evolved stars and in the centers
of spiral and elliptic galaxies. Meanwhile to declare the real
discovery of BHs we need the proof that the suspicious candidates
have a distinctive BH feature --- an event horizon.

In contrast to the evident origin of stellar mass BHs the origin
of SBHs with a mass exceeding $10^6{\rm M}_{\odot}$ in the
galactic nuclei is quite unclear. With a present state of art of
observations and theory we have nowadays only the list of more or
less elaborated scenarios leading to the SBH formation. We will
review here these known ideas and models with discussions of
possible crucial signatures to discriminate between them and find
out the universal mechanism (or several mechanisms ) of SBH
formation. The recent advance and current observational status of
SBH searches is extensively reviewed e.~g. in
\cite{FerrareseFord04,galacticbh}.

The SBHs is not those things that can be ``touched by hands'' or
be investigated thoroughly in the laboratory (it must be noted
that formation of short-lived microscopic BHs is forecasted in
particles collisions at LHC in accordance with some
extra-dimensional theories). Nevertheless, a supposition of their
existence permits to explain a lot of observations. The Seyfert
and radio galaxies, blazars, quasars and huge jets emitted from
the galactic centers could be explained as various aspects of the
same unique phenomenon: an activity of SBH in the galactic center.
It is an interaction of SBHs and matter in the host galaxies that
produces and explains different forms of Active Galactic Nuclei
(AGN) phenomenon \cite{begblrees84,Rees84,Antonucci}.

A recent discovery of very distant quasars reveals the deficiencies
in the understanding of SBH formation. The SBHs are inevitably the
main engine of quasars and the same fact of observation of so
distant SBHs is a source of many questions. Indeed, how could SBHs
as massive as $10^9 - 10^{10}{\rm M}_{\odot}$ be formed so early, at
$z=6-7$, i.~e. at $800$ million years after the Big Bang? After
nucleation, a BH duplicates its mass during the time \cite{Gnedin01}
\begin{equation}
 t_{\rm acc}\simeq4\,10^{7} \mbox{ yrs}, \nonumber
\end{equation}
if an accretion near the Eddington limit takes place. In this case
the smallest period of time necessary to increase an initial BH
mass $M_0$ up to some value $M$ is
\[
t \sim \frac{{4 \cdot 10^7 }}{{\ln 2}}\ln \frac{M}{{M_0 }} \mbox{
yrs}.
\]
Hence about billion years is needed to increase the BH mass from
e.~g. the Solar mass to the observable values. There is no room for
all stages preceding nucleation of the Solar mass BHs --- from a
star formation to its explosion.

Nevertheless, the multi-scale simulations indicate that luminous
quasars at $z \sim 6.5$ could be formed within the framework of
standard $\Lambda$CDM paradigm. It was shown \cite{Li} that the
most massive halos after several mergers could reproduce in
general the observed properties of SDSS J1148+5251, the most
distant quasar detected at $z=6.42$, \cite{Fan}.

\section{Theoretical fundamentals}

The physics of BH is based on the Einstein-Hilbert equations of
general relativity. Specific details and features of BH are
investigated since 1916, when Karl Schwarzschild found the famous
static spherically symmetric solution for a general relativistic
body \cite{Schw}. In spite of complexity of the subject and
permanent disputes on the BH existence, scientific community has
came to agreement on the general BH features. There are a few
important formulas of BH physics which are useful in interpretation
of observational data and astrophysical applications.

A size of BH is characterized by its gravitational radius
\begin{equation}
 r_{g}=
 \frac{2GM}{c^{2}}\simeq
 3\,10^{5}\,\frac{M}{{\rm M}_{\odot}} \mbox{ cm},
 \label{rg}
\end{equation}
where $G$ is the gravitational constant and $M$ is a BH mass.
Another important scale is the minimal radius $r_{\rm st}$ of
stationary circular orbit of a massive body moving around the BH:
\begin{equation}
 r_{\rm st}=3r_{g}
 \label{rst}
\end{equation}
in the case of nonrotating BH.

The SBHs are the most bright objects in the Universe due to
radiation of accreting matter. In a dense environment of galactic
nucleus the mass of SBH is increased in several orders of magnitude
during the lifetime of the galaxy. In the stationary spherically
symmetric case (Bondi accretion) the growth of BH is defined mostly
by the density of surrounded gas. The corresponding rate for BH mass
growth is
\begin{equation}
 \frac{dM_{\rm B}}{dt}=
 4\pi\alpha\frac{G^{2}M^{2}\rho}{c_{s}^{3}},
 \label{Bondi}
\end{equation}
where a dimensionless parameter $\alpha$ depends on the equation
of state of an accreting gas and is of the order of unity, $c_{s}$
is a sound speed of the gas and $\rho$ is its density at infinity.

When a particle is falling onto BH, some part of its energy is
converted into radiation and the rest part increases the mass of BH.
Hence the rate of accretion and the rate of BH mass growth is
proportional to its luminosity. At the same time the accretion
luminosity is limited by the Eddington condition --- equality of the
pressure of radiated photons and gravitational forces experienced by
accreted plasma. This condition leads to the maximal value of
luminosity for the stationary spherically symmetric accretion (the
Eddington luminosity limit):
\begin{equation}
 L_{\rm E}=
 4\pi\frac{GMm_{p}c}{\sigma_{\rm T}}\simeq
 10^{38}\frac{M}{{\rm M}_{\odot}} \mbox{ erg s$^{-1}$},
 \label{LEdd}
\end{equation}
where $m_{p}$ is the proton mass and $\sigma_{\rm T}$ is the Thomson
cross-section.

Equation for BH mass growth in the case of stationary accretion of
baryonic gas is
\begin{equation}
 \frac{dM}{dt}=\frac{M}{t_{\rm Sal}},
 \label{eq1}
\end{equation}
were the Salpeter time $t_{\rm Sal}$ is
\begin{equation}
 t_{\rm Sal}=
 \frac{\varepsilon}{1-\varepsilon}\frac{Mc^{2}}{L}=4.1\,10^{8}
 \frac{\varepsilon}{1-\varepsilon}\frac{L_{\rm E}}{L} \mbox{ yr}.
 \label{tsal}
\end{equation}
Here $\varepsilon$ is an efficiency of accretion and $L$ is an
observable luminosity of BH. It is also reasonable to suppose that
$L\propto M$ as in the Eddington luminosity limit, see
(\ref{LEdd}). In this case the solution of equation (\ref{eq1}) is
rather simple:
\begin{equation}
 M(t)=M_{\rm in}\exp[(t-t_{\rm in})/t_{\rm Sal}].
 \label{Mt}
\end{equation}
where index ``$in$'' corresponds to the initial values.

An accretion efficiency $\varepsilon$ is an emitted fraction of the
total energy of accreted particle. To estimate this value, let us
consider a particle with mass $m$ orbiting around a nonrotating BH
at some radius $r$. If the orbit is stationary, a total energy of
particle is
\begin{equation}
 E=mc^{2}\frac{1-2GM/c^{2}r}{\left(1-3GM/c^{2}r\right)^{1/2}}.
 \label{Er}
\end{equation}
A difference between the energy at last stationary orbit (\ref{rst})
and those at infinity ($r\rightarrow\infty$) is $\Delta
E=0.0572mc^{2}$. Respectively, for the extremely rotating Kerr BH
this value is greater, $\Delta E=0.42mc^{2}$.  A real efficiency
differs from an ``absolute'' value $\varepsilon=0.0572$ for the
Schwarzschild and $\varepsilon=0.42$ for the extreme Kerr case
because some part of radiation is lost within the BH. In addition,
some energy will be radiated during the particle motion between the
horizon and the last stable orbit. For estimations it is usually
used $\varepsilon\simeq0.1$ for an accretion efficiency of
nonrotating BH.

A gravitational redshift $z$ is the productive instrument in modern
astrophysics. Let us denote a difference between the Newtonian
gravitational potentials of two points as $\Delta\Phi_{\rm N}$. The
observers located in these points will measure the different
frequencies $\nu_{1}$ and $\nu_{2}$ of the same electromagnetic
wave. These values define the redshift:
\begin{equation}
 1+z\equiv\frac{\nu_{1}}{\nu_{2}}\simeq
 1+\frac{1}{c^{2}}\Delta\Phi_{\rm N}.
 \label{z}
\end{equation}
An another group of formulae is related to the BH binaries and with
the processes of their collisions and merging accompanied by an
emission of gravitational waves. Investigation of gravitational wave
generation in the Universe is the topical subject in view of coming
completion of huge interferometric laser gravitational waves
detectors. It is known that a collision of two BHs proceeds through
a stage of quasi-periodic motion in close binary after their mutual
gravitational capture. An orbital period for two BHs with mass
$M_{1}$\ and $M_{2}$ is
\begin{equation}
 T_{\rm orb}=2\pi\sqrt{\frac{a^{3}}{G(M_{1}+M_{2})}},
 \label{tbin}
\end{equation}
where $a$ is a semi-major axis of the Keplerian orbit. Obviously,
the gravitational waves emitted by binary have the same period.
Corresponding duration of this Keplerian stage of gravitational
waves emission by binary is \cite{petersmat63}
\begin{align}
 t_{\rm GW}&=
 \frac{5}{256F(e)}\frac{a^{4}c^{5}}{G^{3}(M_{1}+M_{2})^{2}\mu}; \\
 F(e) &  =(1-e^{2})^{7/2}\left( 1+\frac{73}{24}e^{2}+\frac{37}{96}
 e^{4}\right),
 \label{tcoal}
\end{align}
where $e$ is an orbit eccentricity and
$\mu=M_{1}M_{2}/(M_{1}+M_{2})$.

\section{Observational signatures of SBHs}

Strictly speaking we have so far only numerous indirect indications
in favor of BHs existence in the Universe. For this reason the
claimed BHs of stellar mass in binaries and SBHs in galactic nuclei
are only the candidates, not finally proven BHs. There is still some
room for alternatives to BHs among suspected candidates. Decisive
future experiment would be a demonstration or discovery of the BH
event horizon, which is the most striking manifestation of strong
gravitational field and space-time curvature. At the present time,
the crucial point of BH astrophysics is a revealing of observational
signatures of event horizon in suspected BH candidates. The
discovery of a BH event horizon would be the most crucial
verification of the General Relativity as well.

\subsection{Active galactic nuclei and quasars}

Observations of host galaxies of quasars, i.~e. the Quasi Stellar
Objects (QSOs), are rather difficult because of the huge
luminosity of central objects in comparison with stellar
components and because of the large distances. Nevertheless, the
connection between QSOs and their host galaxies has been very
definitely established during the last decade though at least one
exception is known - the bright quasar at the edge of a huge gas
cloud without any visible stellar host galaxy \cite{Mag05}. One of
the explanation consists of a possibility that the SBH powering
the QSO was ejected from the galaxy due to gravitational slingshot
of three or more SBHs in the merger process of smaller galaxies
\cite{HaeDavRee05}. The close similarity of the X-ray spectra of
low- and highest-$z$ QSOs means that SBHs with accretion disks
were already formed $z\geq6$. By using of some type of a
``continuity equation'' with ``source'' in redshift space it is
possible to provide the link between SBHs in distant QSOs and SBHs
in the centers of local galaxies
\cite{CavVit01,haehreeesnar98,KauHae00,choi00}. This analysis is
suitable for a reconstruction of the accretion history of SBHs
\cite{hopkins05,seed06}. Different aspects of the accretion
physics in QSOs are reviewd in \cite{MDS06}.

\subsection{Dynamics of gas and stars near SBH}

Specifical details of motion of gas and stars near the suspected SBH
supply us with valuable information on the gravitational field in
the vicinity of SBH. The best fit of numerous observations gives for
a value of SBH mass at the Galactic Center $M_{\rm BH}\simeq3.6\cdot
10^6{\rm M}_{\odot}$.

Infra-Red (IR) imaging technic allowed to track the Keplerian orbits
of several stars around the Galactic Center. Velocities of the stars
approach $9000$~km~s$^{-1}$ (compare with the average velocity of
stars $200-300$~km~s$^{-1}$). Pericenter distances from the SBH are
as small as 90~AU that is well inside a radius of gravitational
influence of the SBH \cite{Tal05}.

If a star passes close enough to SBH it could be ripped apart by the
tidal gravitational forces. The conditions for such effect are
favorable in the dense stellar nuclei of galaxies.  The flare of UV-
and X-radiation should be emitted by the stream of stellar debris
that plunges into the BH. These star destructions are really visible
phenomena. Gezari et al. \cite{Gez06} observed the flare and its
tail produced by stellar destruction near SBH in the galaxy at
$z=0.37$. Multi-wavelength data collected for 2 years beginning from
the time of the event are in excellent agreement with theoretical
predictions for the tidal disruption of a star. Gas streams from
ripped stars serve the fuel for QSO activity.

On the other hand, a vicinity of SBH is an appropriate area for
the process of young star formation. Well known fact is the
gravitational instability of massive accretion discs \cite{Rice}.
It could lead to formation of new stars if a cooling process in
the disk is effective. Many young stars where discovered in the
nearest vicinity of our Galactic Center, some of them about 0.03
pc from the center \cite{Paumard}. The processes of star formation
and gas accretion could interfere meaningly \cite{Nayakshin}.

In general, a final stage of matter accretion is poorly
investigated yet because of inevitable observational difficulties
in resolving small angular scales. An inner parsec of AGN contains
dense molecular gas of $H_2 O$, $SiO$ and $CH_3 OH$, which is
constantly heated by the accretion generated radiation. A
detection of distribution of water maser emission line at
$1.35$~cm wavelength is a powerful investigation method of the
near vicinity of SBH. A maser activity was first discovered in
1965, \cite{Weaver}. A water maser outside of our Galaxy was
discovered in the spiral arms of the galaxy M33 by
\cite{Churchwell}. Up to now, in total about 60 events are found
\cite{Kondratko}.

Observations of the broad fluorescence $K_{\alpha}$ iron line with
energy $6.4$~keV is also explained in the framework of accreting SBH
in the distant galactic center \cite{Nandra}. Due to the combined
Doppler effect in the fast rotating accretion disk and the
gravitational field of the central BH the observed profile of
emission line has an asymmetric double-peaked shape
\cite{chenhalp,fabreessw,matt,zakharov1}. The fitting of line
profile corresponds to the emission of iron atoms at a distance of
only a few gravitational radius of SBH. From the detailed
spectroscopy of accretion disks it is possible also to evaluate the
Kerr BH spin \cite{fabiwary,zakharov2,pariev1,pariev2}.

New observational evidences continuously confirm the SBH paradigm.
For example, it was reported recently \cite{Shen} the activity of
region close to $SgrA^{\ast}$ at a wavelength $3.5$~mm. This region
surrounds the central BH in the Galaxy having the size about $1$~AU.
On the other hand, the mass of the central object within this scale
is well known, $\sim3\cdot 10^{6}{\rm M}_{\odot}$. It permits to
calculate an average density in the Galactic center, which appears
to be $\sim 6.5\,10^{21}{\rm M}_{\odot}/\mbox{pc}^{3}$
--- too much to consists of gas in any state and/or stars.

\subsection{Galaxy --- SBH correlations}

There are some kinds of observational data which must be explained
by any model of SBH formation. On the other hand, these data give
the ``guiding thread'' to choose the successful models.

Early observations of normal galaxies, quasars and AGN recovered a
simple relation $M_{\rm SBH}\propto L$ between a mass $M_{\rm
SBH}$ of the central SBHs in galactic nuclei and a total
luminosity of host galaxies or their central stellar bulges
\cite{KorRich05,mag98,kazant05}. However an intrinsic data spread
in the above linear relation was very high, even larger then the
observational errors. More tight correlations were found recently
\cite{Gebh} between a mass $M_{\rm SBH}$ and velocity dispersion
$\sigma_e$ at the bulge half-optical-radius:
\begin{equation}
 M_{\rm SBH}\propto\sigma^\alpha,
 \label{korsig}
\end{equation}
where $\alpha=3.75\pm0.3$. An other analysis \cite{Ferra} based on
poorer set of observational data revealed the correlation with a
somewhat different power index $\alpha=4.80\pm0.54$. In \cite{Tre02}
it was shown that difference of slopes $\alpha$ arises mostly due to
systematic difference in the velocity dispersions used by different
authors for the same galaxies. More detailed analysis \cite{Tre02}
of the set of 31 galaxies yields $\alpha=4.02\pm 0.32$. Recently, a
correlation between $M_{\rm SBH}$ and host galaxy velocity
dispersion for QSOs in the Data Release 3 of the Sloan Digital Sky
Survey was discovered \cite{SSGB06}. For small redshifts $z<0.5$
these results agree with ones for the nearby galaxies. For a range
$0.5<z<1.2$ there is indication on the evolution with $z$: the
bulges are too small for their central accreting SBHs. However, this
evolution can be attributed to the observational biases. It have to
be mentioned presumable correlation between the SBH mass and the age
of the galactic stellar population \cite{MerDunForTer02}. This last
type of correlation could be inconsistent with the scenarios of
primordial SBHs origin.

The observed correlations imply one-to-one relation between the
processes of the central SBH and bulge formation and evolution in
the all types of galaxies. Theoretical interpretation of these
correlation is not very easy due to extremely different
length-scales: the galactic bulge scale is a few kpc, whereas the
scale of SBH gravitational influence is millions times smaller.
Specific mechanism is needed for mass transfer from the bulge to its
innermost part. One of the possibilities is that SBH formation
depends on the dynamical properties of bulge through the rate of gas
supply. In this case a velocity dispersion $\sigma$ measures the
depth of the gravitational potential in which a SBH forms and grows.
Another discussed possibility is the energy feedback between an
accreting SBH and bulge which results in a self-regulated accretion
and a corresponding growth of the central SBH \cite{SilRee98}. This
model predicts the following relation, $M_{\rm BH}\propto\sigma^5$.
Third model \cite{HaeKau001} supposes that bulges and SBHs were
formed during the multiple galactic merging. The resulting
correlation between $M_{\rm BH}$ and $\sigma_e$ in the galactic
merging model is in a good agreement with observations. A forth
model proceeds from the hypothesis that observed correlations are
stochastic in origin \cite{DokEro03}. The basic assumption of
stochastic model is an existence of a pre-galactic or primordial SBH
population. A similar hypothesis of an existence of pre-galactic SBH
population was used \cite{FukTur96,DER} for interpretation of the
QSO evolution.

\subsection{Black holes of intermediate mass}

Some mechanisms of SBH formation imply the Intermediate Mass BHs
(IMBHs). The latter are believed to be an intermediate stage of SBHs
growth in astrophysical evolution scenarios with an accretion and
merging. Fast variability of compact X-ray source in the starburst
galaxy M82 indicates that this source is an accreting BH
\cite{Kaa01}. With the observed luminosity of
$10^{41}$~erg~s$^{-1}$, a mass of the supposed BH is of the order of
$\sim10^3{\rm M}_{\odot}$ provided the emission is isotropic and
corresponds to the Eddington limit. The IMBH in M82 exists in
starburst environment. This is a hint on a possible formation of the
IMBH in the M82 by a runaway merging of stelar mass BHs
\cite{MouTan02}. An other suspected IMBH in the star cluster G1 of
the Andromeda galaxy has mass $2\cdot 10^4{\rm M}_{\odot}$
\cite{GebRicHo02}, but there is still a room for a tight cluster of
massive stars in this case.

Some of nearby Globular Clusters may contain IMBHs according to the
measurements of stellar kinematics. On the basis of detailed
observations of Globular Clusters it was stated that IMBHs may
inhabit about a half of Globular Clusters in the Galaxy
\cite{Tre06}. These Globular Clusters are very old in general. In
particular, the mass of IMBH containing in M15 is about $4000{\rm
M}_{\odot}$ \cite{vdM02}. It is interesting that masses of suspected
IMBHs in Globular Clusters are almost those as expected from the
extrapolated $M_{\rm SBH}$-$\sigma$ correlation for galaxies, in
spite of poor statistics of the central BHs with $M_{\rm
SBH}<10^7{\rm M}_{\odot}$ in galactic nuclei.

It was shown \cite{dokero2002} that the primordial IMBHs could be
connected with such dense Dark Matter (DM) objects like neutralino
stars through their common origin from primordial density
perturbations. The neutralino stars with mass $\sim(0.01\div1){\rm
M}_{\odot}$ were proposed to account for observed microlensing
events in the Galactic halo \cite{ufn2}. These neutralino stars
consist of weakly interacting DM particles, and were formed in the
expanding Universe from adiabatic density perturbations. The
existence of neutralino stars should give rise to a large number of
primordial IMBHs with a mass $\sim10^5{\rm M}_{\odot}$. These IMBHs
may constitute a large fraction of the Galactic halo mass.

\subsection{Prospects for gravitational waves detection}

Coalescence of BHs in the galaxies and clusters are inevitably
accompanied by strong bursts of gravitational radiation. The planned
ESA/NASA space observatory LISA will be capable to detect these
coalescence events in very distant galaxies. The combined theories
of galaxies and SBH merges are developed to predict the principally
observed gravitational waves signals
\cite{Haehnelt,MenHaiNar01,WyiLoe02}. Future observations of binary
merging with LISA will allow to test General Relativity in details,
place bounds on alternative theories of gravity and study the merger
history of massive BHs \cite{berti05}.

\subsection{Astrophysical constraints on SBHs}

There are several purely astrophysical constraints on the mass and
number of SBHs. If SBHs provide the dominant part of DM in the
Galaxy, then they must tidally interact and disrupt some part of
Globular Clusters. The SBH mass was constrained in this case,
$M_{\rm{BH}}\le10^4{\rm M}_{\odot}$ \cite{Moore93}. At
$\Omega_{\rm{BH}}\sim1$, primordial BHs are capable of distorting
the CMB spectrum if they are formed about $1$~s after the
annihilation of $e^+e^-$-pairs \cite{carr75}. Mass accretion at the
pre-galactic and present epochs contributes to the background
radiation in different wavelength ranges. However, calculations are
strongly model dependent and yield $\Omega_{\rm{BH}}
\le10^{-3}\div10^{-1}$ for $M_{\rm{BH}}\sim10^5{\rm M}_{\odot}$. In
\cite{nemir}, the constraint on the fraction of intergalactic BHs,
$\Omega_{\rm{BH}}<0.1$, was obtained from a condition of absence of
the reliable gamma-ray-burst lensing events for $10^5{\rm
M}_{\odot}<M_{\rm{BH}}<10^9{\rm M}_{\odot}$. A more stringent
lensing constraint, $\Omega_{\rm{BH}}<0.01$ for the mass range
$10^6{\rm M}_{\odot}<M_{\rm{BH}}<10^8{\rm M}_{\odot}$, was obtained
from VLBI observations of compact radio sources \cite{wilk}.

\subsection{Dark matter spike in the Galactic center}

Annihilation rate of weakly interacting cold DM particles at the
Galactic Center could be greatly enhanced by the growth of density
spike around the central SBH. The resulting annihilation signal may
be observed in the form of high-energy gamma-rays, positrons,
antiprotons and  radio emission \cite{GonSil99}. It could facilitate
the study of DM content. At the same time this annihilation spike
may be the observational signature of the SBH in the Galactic
center. The form and amplitude of this spike is rather uncertain due
to influence of different dynamical processes such as sinking of
baryonic gas and hierarchical merging of sub-halos
\cite{UliZhaKam01,Mer02}. Searching for annihilation signals could
set an upper bound to the DM cusp, DM particle annihilation
cross-section and the merger history of the Galaxy. The DM spikes
could also arise around primordial IMBHs \cite{DokEro03}, and the DM
annihilation in these spikes is in observable range of future
detectors \cite{ZhaSil05}. A model of adiabatic growth of a seed BH
in collisionless DM halo through the DM accretion is proposed in
\cite{MacHen03}. The supposed combined evolution of SBH and DM halo
results in the observed $M_{\rm SBH}-\sigma$ relation.

\section{Astrophysical scenarios of SBH formation}

\subsection{Dynamical evolution of galactic nuclei}

Physically the most simple and natural possibility for SBH
formation is a dissipative dynamical evolution of a dense stellar
cluster in the galactic nucleus ending by its total gravitational
collapse. This idea was first proposed at the dawn of quasar
discovery and then was elaborated in some details in numerous
works.

An evolved galactic nucleus contains dense central stellar cluster
and probably subsystem of compact stellar remnants, such as
neutron stars and stellar mass BHs. This subsystem come from the
dynamical evolution of star cluster in the galactic nucleus
through merging of stars, thereby forming the short-living massive
stars. The clusters of compact remnants form almost inevitably.

To perform estimations, let us consider a central stellar cluster in
the galactic nucleus of mass $M$ and radius $R$ (the radius can be
effectively defined through the sphere which contain half of the
cluster mass), consisting of $N\gg 1$ of identical stars with mass
$m=M/N$ and radius $r_*$. The virial theorem gives the velocity
dispersion in the cluster $v\simeq(GM/2R)^{1/2}$.

A characteristic measure of the stellar system compactness is a
depth of its gravitational potential $|\phi|\simeq v^2$. That is why
the value of star velocity dispersion $v$ is crucial parameter in
evolution history of the stellar system. In the case $v\ll c$ the
Newtonian approximation is applicable to a star motion in cluster.
The Newtonian approximation is failed when a stellar system becomes
relativistically compact, with $|\phi|\simeq c^2$. At this stage of
evolution the radius of the systems is close to its gravitational
radius. The cluster as a whole or its major part becomes dynamically
unstable and collapses into the SBH.

Dynamical evolution of the cluster at nonrelativistic stage was
widely discussed (see e.~g. \cite{spitzer,saslaw,lightshap78,dokrev}
and references therein) and almost exhaustively studied in the
Fokker-Plank approximation. It turns out that very important factors
of the evolution are the tidal interactions of stars, the binary
stars formation and the star mass segregation. Due to the mass
segregation, the binaries and most massive stars are concentrated in
a central part of the cluster.

Dynamical evolution of stellar systems those as galactic nuclei
which are dense enough, proceeds through the stellar collision and
coalescence stage
\cite{ss66,sanders70,spitzerh71,dokrev,MerRep06}. This process is
enhanced by the formation and hardening of star binaries formed by
tidal two-body interactions \cite{fpr75,heggie75,pressteuc77}. The
newly formed hard binaries result in heating of the cluster. The
resulting evolution of star cluster is determined by the
competition of tidal dissipation processes and heating by newly
formes binaries. In the clusters with a high enough velocity
dispersion $v$ dissipative processes dominate and accelerate the
contraction and growth of $v$. Later, due to the growth of
dispersion $v$ the stellar collisions become catastrophic and
destroy the colliding stars. As a result the compact stellar
cluster transforms into the Supermassive Star (SMS) with an
appreciable fraction of neutron stars and and BHs of stellar mass
\cite{begrees78,QuiSha87,QuiSha89,QuiSha90,berdok01,berdok06}.

There are two distinct ways of IMBH or SBH formation in the
evolving star cluster:

(i)  Central core of the cluster reaches relativistic state and
collapses as a whole.

(ii) Stellar mass BHs and neutron stars merge (aggregate),
progressively increasing their mean mass.

In the following we suppose for simplicity that stellar cluster
consists of identical stars of mass $m$.  The main virial parameters
of a self-gravitating stellar system are a total mass $M=Nm$, a mean
radius $R$ within which $N/2$ stars are contained, a star velocity
dispersion is $v$. These parameters obey a virial theorem
\cite{llmech}:
\begin{equation}
 v^2=\frac{GmN}{2R}.
 \label{vvir}
\end{equation}
A total energy of the self-gravitating system is
\begin{equation}
 E=-\frac{1}{2}mNv^2.
 \label{evir}
\end{equation}
These virial relations are established in the dynamical time
$t_{\rm dyn}=R/v$. The corresponding dynamical evolution equation
on the time-scales exceeding $t_{\rm dyn}$ is derived by the
differentiation of viral relations (\ref{vvir}) and (\ref{evir}):
\begin{equation}
 \frac{\dot R}{R}=-\frac{\dot E}{E}+2\,\frac{\dot M}{M}.
 \label{evoleq}
\end{equation}
A central stellar cluster in a galactic nucleus must be compact
enough to evolve during the life-time of the galaxy. Due to the
Coulomb character of the gravitational interactions of stars the
evolution of star orbits in the cluster proceed diffusively with a
characteristic two-body relaxation time \cite{chandra43,spitharm58}
\begin{equation}
 t_{\rm r}=
 \left(\frac{2}{3}\right)^{1/2}\frac{v^3}{3\pi G^2m^2n\Lambda},
 \label{trel}
\end{equation}
where $m$ is a constituent stellar mass, $n$ is a local star number
density, $\Lambda\simeq\log(0.4N)$ is a gravitational Coulomb
logarithm. For many-body stellar systems with number of stars
$N\gg1$, the relaxation time $t_{\rm r}$ is much greater than a
dynamical time $t_{\rm dyn}$:
\begin{equation}
 \frac{t_{\rm r}}{t_{\rm dyn}}
 \sim \frac{N}{\log(N)}\gg1 \quad \mbox{at} \quad N\gg1.
\end{equation}
This means that stars in the self-gravitating dynamically
stationary systems have regular orbits with integrals of motion
(energy and angular momentum) slowly changing at time-scales $t\ll
t_{\rm r}$.

The smallness of the ratio of relaxation time $t_{\rm r}$ and the
age of the galaxy $T_g\sim H^{-1}$ is a measure of stellar cluster
compactness needed for a substantial evolution. Stellar clusters
with $t_{\rm r}/T_g\sim H^{-1}\ll1$ are called (dynamically)
evolved: they have enough time for substantial change of initial
orbital energies and angular momenta of stars in the cluster due
to two-body interactions. The examples of dynamically non-evolved
stellar clusters with $t_{\rm r}/T_g\sim H^{-1}\gg1$: stellar
disks of spiral galaxies, Globular Clusters, galactic DM halos.
The conditions for substantial evolution are realized only in the
central most compact parts of these objects. For the same reason
the central parts of the Globular Clusters and galactic nucleus
are suspected for the final gravitational collapse.

\subsubsection{Non-dissipative contraction of star cluster}
\label{nondis}

A non-dissipative dynamical evolution is governed by the
evaporation of fast stars from the system. The rate of fast star
evaporation due to two-body interactions is
\begin{equation}
 \dot N=-\alpha_{\rm ev}\frac{N}{t_{\rm r}},
 \label{evaprate}
\end{equation}
where numerical coefficient $\alpha_{\rm ev}\simeq10^{-2}$ in the
Fokker-Plank approximation \cite{spitzer,saslaw}. Due to diffusive
nature of star orbital energy changing, an evaporating star escapes
from the stellar system with a nearly zero total energy, i.e. with a
nearly parabolic velocity $v_{\rm esc}=2v$. Hence the total energy
of the stellar system is nearly conserved during star evaporation,
$E\simeq const$. This defines the scaling of the virial parameters:
the evolutionary growing of a star velocity dispersion with a
diminishing of the number of stars, $v\propto N^{-1/2}$, and a
corresponding diminishing of the stellar system radius, $R\propto
N^{-1/2}$. Integration of evolution equation (\ref{evoleq}) at the
condition $E=const$ by using (\ref{evaprate}) and by neglecting
variation of logarithmic factor $\Lambda$ gives a simple law for
non-dissipative dynamical evolution
\begin{equation}
 R(t)=R(0)\left(1-\frac{t}{t_{\rm ev}}\right)^{4/7},
 \label{revap}
\end{equation}
where the lifetime of the system $t_{\rm ev}= (2/7\alpha_{\rm
ev})t_{r}(0)$ and  $t_{r}(0)$ is a relaxation time at the initial
moment $t=0$. According to this formula, a contracting stellar
system reaches the relativistic stage, $v\to c$. At the relativistic
stage, not the evaporation of stars but the dissipative processes
such as gravitational radiation and star collisions become the
crucial dynamical evolution factor. The dissipation processes hasten
the contraction of cluster towards the onset of dynamical
instability and final gravitational collapse. This scenario of
evolution is valid in clusters consisting of degenerate stars like
neutron stars and/or BHs of stellar masses. In the case of
contracting cluster of normal stars, the dissipative tidal
interactions and the star collisions become more important than star
evaporations long before the relativistic stage of evolution.

\subsubsection{Dissipative contraction of star cluster}

The influence of dissipative tidal interactions and star collisions
on the dynamical evolution is crucial in the star clusters with a
large value of stellar velocity dispersion $v$, namely when $v\sim
v_{\rm p}$, where $v_{\rm p}=(2Gm/r_*)^{1/2}$ is a surface parabolic
velocity of constituent stars. For example, in the case of the Sun
$v_{\rm p}\simeq620$~km~s$^{-1}\ll c$.

In close encounters the non-radial oscillations are excited by tidal
interactions of stars. Some part of kinetic orbital energy
transferred into these oscillations. Finally the energy of
non-radial oscillations is thermalized inside stars and reradiated.
This is also the mechanism for dissipative tidal friction of stars
and the tidal capture mechanism of close binary star formation
\cite{fpr75}. The orbital energy deposited into oscillations during
one close encounter equals \cite{pressteuc77}
\begin{equation}
 \Delta E=2^{7/3}\frac{Gm^2}{r_*}
 \left(\frac{r_*}{R_{\min}}\right)^{10}
 \label{deltae}
\end{equation}
for a specific case of stars with $n=3$ polytropic structure. Here
$R_{\min}$ is a periastron distance. The resulting tidal ``cooling''
rate of the stellar system in the case of the Maxwellian
distribution of stars over velocities is
\cite{milgshap78,ozdok82,dokoz82}
\begin{equation}
 \dot E_{\rm dis}=-\Gamma(0.9)g\beta^{-1}\frac{N}{\Lambda t_{\rm r}},
 \quad g=2^{37/30}\left(\frac{Gm^2\beta}{r_*}\right)^{-9/10}.
  \label{endisrate}
\end{equation}
In this equation $\Gamma(0.9)\simeq1.069$ is the Gamma-function and
$\beta^{-1}=mv^2/3$ is a mean kinetic energy of stars in a cluster
(or cluster's ``temperature''). The tidal dissipative friction of
stars in the cluster accelerates its contraction. Under the
influence of both evaporation of stars and tidal friction the
evolution equation (\ref{evoleq}) according to (\ref{evaprate}) and
(\ref{endisrate}) takes the form
\begin{equation}
 \frac{\dot R}{R}=
 -\left[\left(\frac{v}{v_{\rm dis}}\right)^{9/5}
 +2\right]\,\frac{\alpha_{\rm ev}}{t_{\rm r}},
 \label{disevoleq}
\end{equation}
where
\begin{equation}
 v_{\rm dis}=\left(\frac{3}{2}\right)^{1/2}2^{-67/54}
 \left[\frac{3(\pi/2)^{1/2}
 \alpha_{\rm ev}\Lambda}{\Gamma(0.9)}\right]^{5/9}v_{\rm p}.
 \label{vdis}
\end{equation}
For a case of cluster consisting of the Sun type stars, $v_{\rm
dis}\simeq190$~km~$s^{-1}$. According to equation (\ref{disevoleq})
dynamical evolution follows non-dissipative evaporative scenario
(\ref{revap}) until $v\ll v_{\rm dis}$. It also concerns the central
parts of the nowadays Globular clusters with $v\sim10$~km~$s^{-1}$.
Only a very small fraction of stars in Globular Clusters survives in
the final steps of dynamical evolution. Additionally a lot of hard
binaries (with binding energies exceeding $\beta^{-1}$) are formed
in Globular Clusters by dissipative two-body encounters
\cite{fpr75}. As a result the heating of cluster by these hard
binaries \cite{heggie75} prevents the formation of a massive BHs in
the Globular Clusters (see for details \cite{ozdok82,dokoz82}).

\begin{figure}[ptb]
\begin{center}
\includegraphics[width=0.95\textwidth]{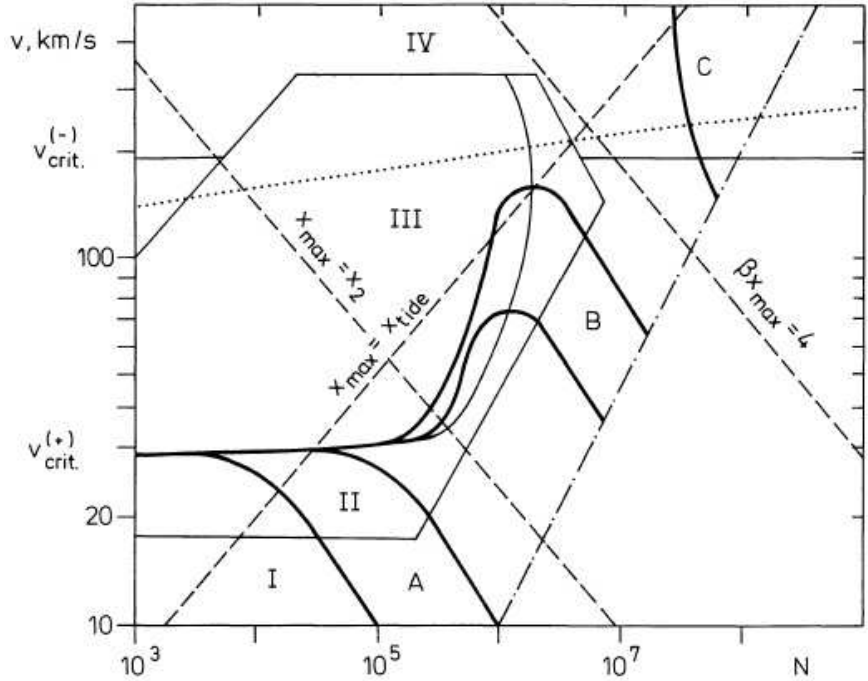}
\end{center}
\caption{Evolution tracks of stellar systems in the plane ``number
of stars $N$, velocity dispersion $v$'' (see \cite{dokoz82} for
details). $A$ --- globular clusters, $B$ --- normal galaxies with
low $v$, $C$ --- galactic nuclei with high $v$ In region I, the
tidal dissipative effects and heating of stellar system by newly
produced binaries do not influence the evolution of the system which
follows to the non-dissipative evaporative scenario (see
Section~\ref{nondis}). The boundary between regions I and II (and
between III and IV) corresponds to condition $\dot E=0$, when
dissipative effects in the system are compensated by the heating by
binaries. In the region II, the heating of the system by newly
formed binaries dominates over the tidal dissipation, $\dot E>0$.
The boundary between regions II and III is determined by the
condition $\dot v=0$, or equivalently $v=v_{\rm crit}^{(+)}$. In the
region III, where $v>v_{\rm crit}^{(-)}\sim190$~km~s$^{-1}$, the
heating by binaries prevails over the tidal dissipation and $v$
decreases with the decrease of $N$. In the region IV, the evolution
of stellar system is mainly affected by tidal dissipative energy
losses which lead to a rapid contraction of the system toward to the
final collapse into SBH. Above the dotted line the rate of star
collisions becomes greater than the rate of star evaporation. The
dashed-dotted line corresponds to $t_{\rm r}=10^{10}$~yr.}
 \label{dissipevol}
\end{figure}

On the contrary, in stellar systems such as the galactic nuclei
with $v\geq v_{\rm dis}$, the tidal friction dominates over star
evaporation and heating by hard binaries. As a result the cluster
contracts formally to a zero radius having a finite number of
stars in the remnant \cite{milgshap78,dokoz82}. Physically this
dissipative stage of evolution is finished by the direct
collisions, merging and destruction of stars. Collisions of stars
with relative velocity $u<v_{\rm p}$ result in merging of stars or
their explosion of a supernova type. A head-on collisions of stars
with $u>v_{\rm p}$ are more dangerous and inevitably end with a
total destruction of these stars \cite{ss66}. In this limit the
rate of disruptive collisions of stars in the system is equal to
\begin{equation}
 \dot N=
 \left(\frac{v}{v_{\rm p}}\right)^{4}\frac{N}{\Lambda t_{\rm r}}.
 \label{collrate}
\end{equation}
A contracting star system with $v>v_{\rm p}$ consists of a
decreasing number of stars and an increasing number of moving gas
clouds --- remnants of destructive star collisions. These gas clouds
participate in the virial balance as the peer entities until they
are small compared with the size of the remaining system. In a
simple approximation, suppose that the two clouds with a nearly
equal masses are formed as a result of the two stars catastrophic
collision, and so $Nm=M=const$. Integration of equations
(\ref{evoleq}) and (\ref{collrate}) in this approximation gives
\begin{equation}
 R(t)=R(0)\left(1-\frac{t}{t_{\rm coll}}\right)^{2/7},
 \label{evolcoll}
\end{equation}
where the collision evolution time $t_{\rm coll}=[v_{\rm
p}/v(0)]^4\Lambda t_{\rm r}(0)$. The system contracts according to
(\ref{evolcoll}) until its radius reaches $R\sim N^{1/2}r_*$, at
which point $t_{\rm coll}\sim t_{\rm dyn}$. At this point the system
ceases to exist as a collection of separate stars and transforms
into the compact self-gravitating massive cloud of gas
--- the Supermassive Star (SMS)
\cite{hoylefowler63,fowler66,fowler66b,zeldnov71}. The possible
evolution tracks of stellar clusters are shown in the
Fig.~\ref{dissipevol}.

\subsection{Collapse of supermassive star}

A Supermassive Star (SMS) was first proposed long ago as a possible
source of quasar activity and nowadays is considered as a short
intermediate stage of galactic nucleus evolution toward the
formation of SBH. To prevent a fragmentation before a SMS formation,
a primordial gas must be hot and magnetized.
\cite{haehreeesnar98,haehrees93,eisloeb95}. The natural way for SMS
production is stellar collisions stage of evolution of compact
galactic nuclei \cite{sanders70,begrees78,begblrees84,Rees84}.

Fast evolving SMSs include an unstable hydrodynamic relativistic
radial mode
\cite{fowler66,fowler66b,iben63,chandra64,chandra64b,chandra65} and
eventually collapse to form one or several close SBHs
\cite{fowler66,fowler66b,zeldnov71,lee87}. Recently numerical
simulations provides the possibility to analyze the final state of
the SMS gravitational collapse
\cite{sbss02,shibshapbss02,shibata03}.

The SMSs may be naturally formed in the galactic nuclei from gas
produced in the destructive collisions of stars in the evolved
stellar clusters with a velocity dispersion $v\geq v_{\rm p}$.
Characteristic time-scale for the dynamical evolution of stellar
cluster is the (two-body) relaxation time $t_{\rm r}$. According to
(\ref{trel}) it can be expressed as
\begin{equation}
 \label{tr}
 t_{r}\simeq
 4.6\times10^8N_8^2\left(\frac{v}{v_{\rm p}}\right)^{-3}
 \!\mbox{ yr},
\end{equation}
where $N=10^8N_8$ is the number of stars in the cluster. The
corresponding virial radius of the stellar cluster is $R=GNm/2v^2$.
At $v>v_{\rm p}$, where $v_{\rm p}$ is an escape velocity from the
surface of constituent star, the time-scale for self-destruction of
normal stars in mutual collisions is $t_{\rm coll}=(v_{\rm
p}/v)^4\Lambda t_{\rm r}$ \cite{dokrev}. Numerical modelling of
catastrophic stellar collisions has been performed by e.~g.
\cite{ben92,lai93}. We choose $v\simeq v_{\rm p}$ as a
characteristic threshold value for a complete destruction of two
stars and final production of unbound gas cloud. As a result, the
stellar cluster in the evolved galactic nucleus with $v\geq v_{\rm
p}$ transforms into the SMS due to catastrophic stellar collisions.
At $v\simeq v_{\rm p}$ the time-scale for the formation of SMS due
to destructive collisions of stars is of the same order as the
relaxation time, $t_{\rm coll}(v_{\rm p})\sim t_{\rm r}(v_{\rm p})$.
A total mass of the gas produced by destruction of normal stars
composes the major part of a progenitor central stellar cluster in
the galactic nucleus. The natural range of masses for the formed SMS
is $M_{\rm SMS}=10^7-10^8{\rm {\rm M}_{\odot}}$.

A newly formed SMS with mass $M_{\rm SMS}$ and radius $R_{\rm SMS}$
gradually contracts due to radiation with the Eddington luminosity,
$L=L_{\rm E}$. A nonrotating SMS is a short-lived object that
collapses due to post-Newtonian instability. Rotation provides the
stabilization of SMS if the rotation energy is an appreciable part
of its total energy $E_{\rm SMS}\simeq(GM_{\rm SMS}^2/2R_{\rm
SMS})$. In general an evolution time of SMS is the Kelvin-Helmholtz
time-scale $t_{\rm SMS}= E_{\rm SMS}/L_{\rm E}\propto R_{\rm
SMS}^{-1}$. Stabilization of SMS by rotation (and additionally by
internal magnetic field) ensures in principle its gradual
contraction up to the gravitational radius
\cite{zeldnov71,shapteuk83,new01}. A resulting maximum evolution
time of SMS is of the order of the Eddington time $t_{\rm
E}=0.1M_{\rm SMS}c^2/L_{\rm E}\simeq4.5\times10^7$~yr. The
distribution of gas in a SMS can be approximated by the polytropic
model with an adiabatic index  $\gamma=4/3$. For this value of
adiabatic index the central gas density in SMS is $\rho_{c}=k_{\rm
c} n_{\rm SMS}m_{\rm p}$ and central sound velocity $c_{\rm
s,c}=k_{\rm s} v_{\rm SMS}$, where $n_{\rm SMS}$ is a SMS mean
number density, $v_{\rm SMS}= (GM_{\rm SMS}/2R_{\rm SMS})^{1/2}$ is
a SMS virial velocity and numerical constants $k_{\rm c}\simeq54.2$
and $k_{\rm s}\simeq1.51$ respectively.

A SMS, formed in the galactic nucleus, may contain compact remnants
of exhausted stars in the form of neutron stars and stellar mass
BHs. Due to dynamical friction these compact objects are settled
down to the central part of the SMS. They form a very compact and
fast evolving self-gravitating subsystem which collapses into the
massive BH earlier than the host SMS. An observational signature of
these nearly collapsing SMS in the distant galactic nuclei may be a
long-wavelength gravitational radiation \cite{kimbshap01} and a
power flux of high-energy neutrino \cite{berdok01,berdok06}.

\subsection{Collapse of neutron star cluster}

Neutron star cluster evolves much faster than a host SMS and
collapses finally into the massive BH. The dense clusters of neutron
stars and stellar-mass BHs can undergo the catastrophic events of
gravitational collapse due to general relativistic instability. The
collapse goes through the avalanche-type contraction described
firstly in \cite{ZelPod65} and confirmed by the detailed numerical
calculations of
\cite{QuiSha87,QuiSha89,QuiSha90,ShaTeu,ShaTeu2,ShaTeu3,ShaTeu4}.
They found that even in the case of extremely centrally condensed
cluster configuration with an extensive Newtonian halo, a
significant fraction (several percents) of total mass collapses into
a central BH in a few dynamical times. The collapse begins when the
central redshift approaches $z\simeq 0.5$ and a corresponding virial
radius of the cluster diminishes up to $R\sim3 R_g$. The `avalanche'
arises because the falling of stars onto the center leads to an
increase of the gravitational field. The latter acts on the other
particles/stars, whose orbits contract in turn, and so on. The
non-circularity of star orbits are of principle significance because
they connect the different layers of the cluster. Only stars at
highly elongated orbits are involved in the collapse and accreted
into the growing BH. After the collapse the cluster settle down to a
new, dynamically stable equilibrium state: central BH embedded into
quasi-Newtonian cluster of stars and compact stellar remnants. This
evolutionary track is time-consuming because it requires the prior
evolution of normal stars and subsequent settle down of their
remnants to a highly concentrated system.

The further growth of a seed massive BH with a mass $M_{\rm
BH}\geq10^4{\rm M}_{\odot}$ is governed by an accretion of ambient
stars, gas and DM particles (see e.~g.
\cite{begblrees84,dokrev2,ilyin04}). During a lifetime of the
normal galaxy a seed massive BH should grow significantly, up to
$M_{\rm BH}\sim10^6-10^8{\rm M}_{\odot}$.

\subsection{Merging of stellar-mass black holes}

The alternative model of IMBH formation in a star cluster is the
multiple merging of stellar mass BHs when. The evolution of mass
distribution usually studied by using the Smoluchowski equation
\cite{MouTan02,benson05,erickcek06}. Under the conditions that may
exist in some star clusters the characteristic timescale for runaway
merging is $\sim10^7$~yr, which is short enough for the formation of
BH with mass $\sim10^3{\rm M}_{\odot}$. More massive BHs sink into
the cluster center under the influence of dynamical friction and
increase the merger probability in the core. Such a 'mass
segregation' is very important process for the evolution of the
cluster core. After merging of hosted protogalaxies, the IMBHs may
sink to the galactic centers and merge into SBHs. It should be
mentioned that this evolutionary track is rather time-consuming and
probably can not explain the observed early QSO activity. Additional
prolific accretion of baryons or DM is required \cite{TylJanSan03}.
It is considered also a possibility of the super-Eddington accretion
regime, \cite{Rees92,Kawaguchi} which strongly accelerates a BH
growth.

A peculiar example is an accretion onto BHs of the Dark Energy (DE)
with an equation of state $P(\rho)<0$. In the extreme case of
phantom energy (with $P+\rho<0$), the accretion is accompanied with
a gradual decrease (rather than increase, as in a case of the usual
matter) of the BH mass \cite{bde03,bde03b}. Respectively, masses of
all BHs tend to zero while the phantom energy Universe approaching
to the Big Rip.

\subsection{Merging trees of galaxies and SBHs}

Galaxies are formed by hierarchical merging of smaller
protogalaxies, and even nowadays a significant fraction of low
redshift galaxies still experience merging. When galaxies merge,
their cental SBHs get a chance to merge too. The dynamical friction
in the field of stars and DM brings the SBHs into the central part
of coalesced galaxies, where they finally merged into a bigger SBH.
The timescale for a $10^8{\rm M}_{\odot}$ SBH to spiral down into
the nucleus of typical giant galaxy from an initial radius of
$10$~kpc is $\sim10^{10}$~yr. A subsequent closing of two SBHs to
one another proceeds even faster $\sim10^7$~yr due to a high number
density of stars in the galactic nucleus, which produces a strong
dynamical friction \cite{FukEbiMak92,HaeKau02}. On the final stage
of merging the gravitational radiation dominates. However, there are
calculations which predict the long living binary SBHs, longer than
the Hubble time \cite{ValVal89}.

Merging of small protogalaxies containing the IMBHs is a promising
mechanism of SBH formation. The seed IMBH could be formed
initially in the mini-haloes collapsing at $z\sim 20$ from
high-$\sigma$ density fluctuations
\cite{VolHaaMad02,BroLoe02,IslTaySil03}. They increase their
masses up to modern values $10^6-10^9{\rm M}_{\odot}$ due to
multiple merging and accretion \cite{cattaneo05,VolLodNat07}.
Direct evidence of SBHs merging was founded by VLA telescope from
observation of jets flipping (X-shape of radio lobes) in NGC~326
galaxy \cite{merritt02}. It is interesting that a total energy
transferred from the binary SBH to the field stars is comparable
with a binding energy of galactic core \cite{FukEbiMak92}. A
combined upper limit on the mutual evolution of SBHs and their
host galaxies was obtained in \cite{HRKHC06}.

Recent X-ray observations clearly demonstrate the existence of
binary AGN in the galactic system Arp~299 consisting of two galaxies
in an advanced merging state, NGC~3690 and IC~694 to the east, plus
a small compact galaxy \cite{ballo03}. The other cases of two AGN in
a merging system is NGC~6240 \cite{komossa03} and NGC~6104
\cite{smirnova06}. A close encounter and coalescence of spiral
galaxies triggers an inflow of nuclear gas, which fuels both a
powerful starburst and strong accretion onto the central SBH
\cite{springel05,escala05}.

Coalescence of SBHs in galaxies must be accompanied by the strong
bursts of gravitational radiation. The planned ESA/NASA space
observatory LISA will be capable to detect these coalescences and
verify model predictions for SBHs merging rate
\cite{merritt02,campan05}. A binary SBH leaves an imprint on a
galactic nucleus in the form of a ``mass deficit'', a decrease in
the mass of the nucleus due to ejection of stars by the binary
\cite{merr06}. Comparison with observed mass deficits implies
between 1 and 3 mergers for most galaxies, in accordance with the
hierarchical galaxy formation models.

A mean number of galactic mergers is rather low to ensure the
growth of stellar mass BHs to SBHs in most galaxies with the only
exception of giant CD galaxies in the centers of rich clusters.
Another difficulty is the ``final parsec problem'': quite large
distance between SBHs in binaries after the merging of host
galaxies to provide the strong enough gravitational wave radiation
losses for merging of SBHs in binaries \cite{berczik06}.
Nevertheless, it seems that SBHs populate near the all spiral and
elliptical galaxies. This may be a hint in favor of some universal
mechanism of SBH formation long before the formation of galaxies
in the early Universe. The merging rate of galaxies and their
central SBHs is insufficient to ensure the needed mass growth at
least at low redshifts, $z<0.36$ \cite{masjedi06}.

\section{Cosmological scenarios of SBH formation}

The present state of art in understanding of SBH origin provides
only a list of different scenarios, more or less elaborated. Even
Moreover, we even non confident in the choosing between two major
alternatives: the galactic or the cosmological origin of these
objects. In view of these uncertainties we will itemize in this
review the most popular known scenarios with a brief elucidation of
underlying physics. It is a challenge for future theoretical
investigations and detailed astrophysical observations to recover
the real mechanism(s) of SBH formation.

The problem of SBH origin becomes more acute with a discovery of
very distant quasars. Most distant of them are:
\begin{align}
 J114816.64+525150.3,\quad z &
 =6.43,M\simeq3\,10^{9}{\rm M}_{\odot}; \nonumber \\
 J103027.10+052455.0,\quad z &
 =6.28,M\simeq3.6\,10^{9}{\rm M}_{\odot}; \\
 J130608.26+035626.3,\quad z  &
 = 5.99,M\simeq2.4\,10^{9}{\rm M}_{\odot}. \nonumber
\end{align}
Observations of high redshift quasars lead to an unambiguous
conclusion that host galaxies of SBHs are formed very early. The
existence of quasar with redshift $z=6.43$ indicates that SBHs with
mass around $10^{9}{\rm M}_{\odot}$ already existed when the
Universe was less than 1~Gyr old. In order of magnitude, it is just
a lifetime of an ordinary massive star. Such stars produce BHs with
masses of order $1{\rm M}_{\odot}$ after explosion. Using formula
(\ref{Mt}) one can obtain the time needed a BH to grow up in
$10^{9}$ times. In case of the Eddington accretion, a BH increases
its mass in $10^{9}$ times during $800$ million years. There is no
room for first ordinary star formation.

Nowadays we have a list of competing models to solve this puzzle.
This list consists of two parts: cosmological (pre-galactic)
astrophysical (galactic) models. The latter were discussed in
previous Section. Before the discussion of cosmological models let
us consider a couple of ``intermediate'' models where SBHs and
galaxies are formed at the same time approximately.

\subsection{Population III stars or supermassive stars}

BHs with masses about $100{\rm M}_{\odot}$ are thought to be
produced as remnants of first stars (Population III stars) which are
supposed as massive as $200{\rm M}_{\odot}$
\cite{Madau2001,All2,All2b,All2c}. A lifetime of these stars is
quite short --- after several millions years they collapse into BHs.

According to a hierarchical model of structure formation, less
massive DM objects are formed earlier. Formation of gravitationally
bound objects with $\sim10^5-10^7{\rm M}_{\odot}$ is expected to be
before $z\sim10$. The short-lived massive stars with mass
$\sim100{\rm M}_{\odot}$ can appear from the metal-free gas clouds
in the first DM sub-haloes, with a several massive stars per a DM
sub-halo.

Formation of population III stars requires an effective cooling
mechanisms at early epoch of fragmentation of primeval gas clouds
into the stars. Usually believed that the line emission of molecular
hydrogen is the basic cooler in these primeval gas clouds. If the
gas cloud evolves without fragmentation to stars, it can reach the
state of a single Super-Massive  Star (SMS) with mass in the range
$\sim10^3-10^5{\rm M}_{\odot}$. A SMS is a very short-lived object
and inevitably collapses into an IMBH.

\subsection{Collapse of massive primordial clouds}

SBHs could be formed as a result of dissipation and collapse of
primordial gas during the early stages of galaxy formation
\cite{Gnedin01,haehreeesnar98,KouBulDek04,All1,All1b,All1c,All1d}.
Initial masses of such SBHs are of order $10^{5}{\rm M}_{\odot}$.
Another scenario of SBH formation in the low angular momentum
haloes was proposed and recently considered in details by
\cite{KouBulDek04}. In this scenario the baryons cool and sink to
the center of the halo and finally settle down into a rotating
disk. Due to viscosity, an angular momentum is transferred outward
from an inner region of the disk. A model of the formation of
lightweight BHs (a few solar mass) by the direct collapse of
gaseous clouds (without a preceding formation of stars) was
elaborated in \cite{Beg07}. It is supposed that these BHs are
growing up to the supermassive ones by accretion of massive
envelops with the super-Eddington rate.

The haloes are distributed over their angular momenta. This
distribution depends only on the spectrum of primordial density
perturbations and was obtained from analytical calculations and
numerical simulations
\cite{SusSasTan94,Loe93,LoeRas94,EisLoe95,EisLoe95-2}. Only those
haloes from low angular momentum tail of the distribution are
capable to collapse. According to \cite{KouBulDek04}, haloes more
massive than a critical threshold about $\sim 7\cdot 10^7{\rm
M}_{\odot}$ at red shifts $z\sim 15$ contain the gravitationally
unstable discs with an efficient viscosity. The authors of
\cite{KouBulDek04} also showed that under reasonable assumptions,
this model leads to the observed correlation between BH masses and
spheroid properties.

Combined dynamics of baryons and DM in the halo of forming galaxy
was analyzed in \cite{GurZyb90}. The cooling and contraction of
baryonic component both lead to the formation of SBH which grows
rapidly due to accretion as baryons \cite{SIZG04} and the DM
\cite{IZG04,ZelVas05}. Absorption of DM in \cite{IZG04} and
\cite{ZelVas05} was examined by taking into account the scattering
of DM particles on stars near the galactic center. It was shown that
significant flux of DM on a seed BH in the galactic center of galaxy
dominates. Respectively, a fraction of DM in the total mass of the
SBH in this scenario is significant.

\subsection{Pre-galactic SBH}
As it was mentioned above there were proposed some mechanisms of
massive BH production {\em before} the formation of first stars.
Below we shortly list the proposed mechanisms of primordial BH
formation (see also the detailed review \cite{Carr}) and discuss
some of the listed models in more details in the next sections.

\textbf{Adiabatic fluctuations at radiation-domination stage}.
According to \cite{carr75,zeldnov66,hawking71,Khlopov80}, density
fluctuations at the radiation-dominated evolutionary stage of the
Universe give rise to primordial BHs. The formation of BHs is
quite efficient in the case of ``blue'' spectra of power-law
density perturbations $n\ge1.1-1.2$.

\textbf{Isocurvature (or isothermal) fluctuations after
inflationary stage}. Variety of models with the flat or bumped
spectra of isocurvature fluctuations were proposed
\cite{Barrow81,Bond82,Fuk86,Dolg87,Kof87,Kof88,Dol91,CDol92,Iok91}
A very promising mechanism of SMB formation from the large
amplitude isothermal fluctuations in baryonic charge density was
proposed by Dolgov and Silk \cite{DolgovSilk93}. In this mechanism
the bumped spectrum of isothermal fluctuations is a byproduct of
vacuum bubble collapses during phase transition at the
inflationary stage.

\textbf{Artificial form of the inflaton potential}. Models of such
sort represent a modification of the previous model. It is known
that the inflationary stage is the reason of density fluctuations.
Spectrum of the fluctuations depends on the form of the inflaton
potential, see i.~g. \cite{Star92,ivan94}. As it was shown in
\cite{Yoko}, it opens a possibility to produce high density
fluctuations that collapse into SBHs in the following.

\textbf{SBHs from phase transitions}. BHs with masses $\sim1{\rm
M}_{\odot}$ possibly formed at quark-hadron phase transition at the
cosmological moment $10^{-6}$~s, \cite{Jedamzik,Jedamzik2}. Such BHs
would be a component of DM today.

\textbf{Soft equation of state}. Suppose that some massive
non-relativistic particles dominates at an early period of the
Universe evolution. In that case the pressure is negligible and
could not prevent gravitational collapse of high density regions
\cite{Khlopov80}.

\textbf{Bubble collision}. False vacuum decay is usually accompanied
by formation of spherical walls with a true vacuum inside them
\cite{Bubble,Bubble2,Bubble3,Bubble4,Bubble5,Bubble6,Bubble7,Ru25}.
The walls quickly expand and collide what could lead to collapse of
islands of false vacuum into BH. Most massive BHs of order $1{\rm
M}_{\odot}$ are produced during the period of quark-hadron phase
transition.

\textbf{Closed domain walls}. A mechanism of BH formation from the
closed domain walls was proposed in \cite{Ru1,Ru01b,Ru05}. These
domain walls could be originated due to evolution of a scalar field
during inflation. An initial non-equilibrium distribution of scalar
field imposed by the background de-Sitter fluctuations gives rise to
the spectrum of BHs, which covers a wide range of mass --- from the
subsolar up to supermassive ones. The primordial BHs of smaller
masses are concentrated around the most massive ones within a
fractal-like cluster. It was revealed that this mechanism is a
rather common for many inflationary models and it worth to discuss
its essential details at the end of the review.

\subsection{Adiabatic fluctuations at radiation-domination stage}

Gravitational collapse and the formation of a primordial BH occurs
if a relative radiation density fluctuation $\delta_{\rm{H}}=
(\rho-\rho_{\rm{c}})/\rho_{\rm{c}}$ satisfies some conditions
\cite{carr75,zeldnov66,hawking71} at the moment $t$ of its
separation from the cosmological expansion. Namely, the BH is
nucleated if
\begin{equation}
 \delta_{\rm{c}}\le\delta_{\rm{H}}\le1,
 \label{usl}
\end{equation}
where $\delta_{\rm{c}}=1/3$ \cite{carr75}. The left-hand inequality
implies that the radius of the perturbed region at the time $t$
exceeds the Jeans radius $ct/\sqrt{3}$. The right-hand inequality
corresponds to the formation of a primordial BH rather than an
isolated Universe. The primordial BHs are produced with a near
horizon size, and thus their masses are connected with the formation
time as
\begin{equation}
 t=\frac{GM}{c^3}=
 0.5\left(\frac{M}{10^5{\rm M}_{\odot}}\right)\mbox{~s}.
 \label{th}
\end{equation}
In recent years, numerical experiments have revealed a near-critical
gravitational collapse of primordial perturbations with less than
the horizon scales \cite{niem98,niem99,niem99b,chop}. In this
process a mass of forming primordial BH is
\begin{equation}
 M_{\rm{BH}}=
 AM_{\rm{H}}(\delta_{\rm{H}}-\delta_{\rm{c}})^{\gamma},
 \label{critcol}
\end{equation}
where $A\sim 3$, $\gamma\simeq0.36$, and
$\delta_{\rm{c}}\simeq(0.65\div0.7)$. The mass (\ref{critcol}) may
be much smaller than the horizon mass $M_{\rm{H}}$. However, as
shown in \cite{yok98}, the primordial BH mass distribution for
critical gravitational collapse is concentrated near
$M_{\rm{BH}}\sim M_{\rm{H}}$, and the contribution of low masses to
the cosmological primordial BH density is modest.

If the power spectrum of primordial cosmological perturbations is a
power law with an index $n>1$, then the primordial BHs are formed in
a wide range of masses. However a power law spectrum with $n>1$ is
not favored by recent CMB observations, and a corresponding number
density of resulting primordial BHs is extremely (exponentially!)
small. This is because the primordial BHs form at the tail of the
Gaussian distribution with probability
\begin{equation}
 \beta=
 \int\limits_{\delta_{\rm{c}}}^{1} \frac{\displaystyle
 d\delta_{\rm{H}}}{\displaystyle\sqrt{2\pi}\Delta_
 {\rm{H}}} \exp(-\frac{
 \displaystyle\delta_{\rm{H}}^2}{\displaystyle2\Delta_{\rm
 {H}}^2})\simeq\frac{\Delta_{\rm{H}}}{\delta_{\rm{c}}
 \sqrt{2\pi}}\exp(-\frac{\delta_
 {\rm{c}}^2}{2\Delta_{\rm{H}}^2}).
 \label{bet2}
\end{equation}
The value of r.m.s. fluctuation at the horizon scale
$\Delta_{\rm{H}}$ can't reach value $\simeq0.05$ which is necessary
to achieve $\beta\sim10^{-8}$ and correspondingly the cosmological
mass fraction of primordial BHs $\sim10^{-3}$ today.

If, however, the spectrum has a peak at some scale, then primordial
BHs are formed mostly in a narrow range of masses, near the mass
that corresponds to this peak. The spectrum with a sharp maximum at
some scale arises in some inflation models if the inflationary
potential $V(\phi)$ has a flat segment (see e.~g.
\cite{Star92,ivan94}). As it was shown in \cite{Yoko}, the strong
density fluctuations are generated in this models and collapse into
SBHs. At the same time the spectrum beyond the maximum may be of the
standard Harrison-Zeldovich form and reproduce the usual scenario of
large-scale structure formation in the galactic distribution. The
primordial BHs production from the maximum may be connected to the
formation of noncompact DM objects at the matter-dominated epoch
\cite{DokEro01}.

For the early formed primordial BHs, the process of mass increasing
by the secondary accretion of DM is possible. Indeed, in the uniform
Universe a primordial BH with mass $M_h$ inside the sphere
containing total mass $M$ produces the gravitational fluctuation
$\delta_{\rm eq}=M_h/M$ at the epoch of matter-radiation equality
$t_{\rm eq}$. Subsequent growth of this fluctuation obeys the
well-known law for DM fluctuations. As a result the primordial BHs
would be ``dressed'' by the DM halo with a steep density profile,
$\rho\propto r^{-9/4}$. We call this combined spherical volume of
the ``primordial BH~$+$halo'' by the ``induced halo''. It was shown
\cite{DokEro03} that for primordial BH with a mass $\sim10^5{\rm
M}_{\odot}$ the resultant mass of induced halo is of the order of
mass of a typical dwarf galaxy, $\sim10^7{\rm M}_{\odot}$. During a
hierarchical clustering of DM the induced halos enter into the
haloes of ordinary galaxies. Many of these induced halos are capable
to sink down to the galactic center during the Hubble time under the
influence of dynamical friction.

\subsection{Clusters of BHs from closed domain walls}

\subsubsection{General idea}

Some inflationary models suppose a creation of our Universe either
near a maximum of potential of inflaton field or near its saddle
point(s) to realize a desired slow rolling providing a sufficient
number of e-foldings (see details, e.~g. in \cite{Dvali,Racetrack}).
As it will be shown below these models include the possibility of
the formation of macroscopically large closed walls from a scalar
field. After the end of inflation these closed walls collapse to BHs
if these walls are large and heavy enough \cite{Ru1,Ru01b}. This
mechanism is realized in well known models like the Hybrid Inflation
\cite{LindeHyb} and the Natural Inflation \cite{Dolgov97}. A scalar
field could be the inflaton itself or some additional field.

First of all we consider a general mechanism of closed wall
formation based on the quantum fluctuations near unstable point(s)
like a saddle point or a maximum of potential of scalar field. An
evolving scalar field may be split into a classical part, governed
by the classical equation of motion, and the quantum fluctuations
\cite{Star}. To facilitate the analysis, let us approximate a
potential near its maximum as
\begin{equation}
 V = V_{0} - \frac{m^{2}}{2} \phi^{2},
 \label{Vapprox}
\end{equation}
where the maximum is assumed at $\phi=0$ without the loss of
generality. Then, a probability density to find a certain field
value $\phi$ has a form \cite{Book1} (adapted to the considered
case):
\begin{equation}
 dP(\phi,T;\phi_{\rm{in}},0)=
 d\phi\sqrt{\frac{a}{\pi(e^{2\mu T}-1) }}
 \exp\left[-a\frac{(\phi-\phi_{\rm{in}}e^{\mu T})^{2}}
 {e^{2\mu T}-1}\right].
\end{equation}
Here $a=\mu/\sigma^{2}$, $\mu\equiv m^{2}/3H$ and $\sigma=
H^{3/2}/2\pi$, where the Hubble parameter $H\simeq\sqrt{8\pi
V_{0}/(3M_{\rm{Pl}})}$.

\begin{figure}[ptb]
\begin{center}
\includegraphics[width=1\textwidth]{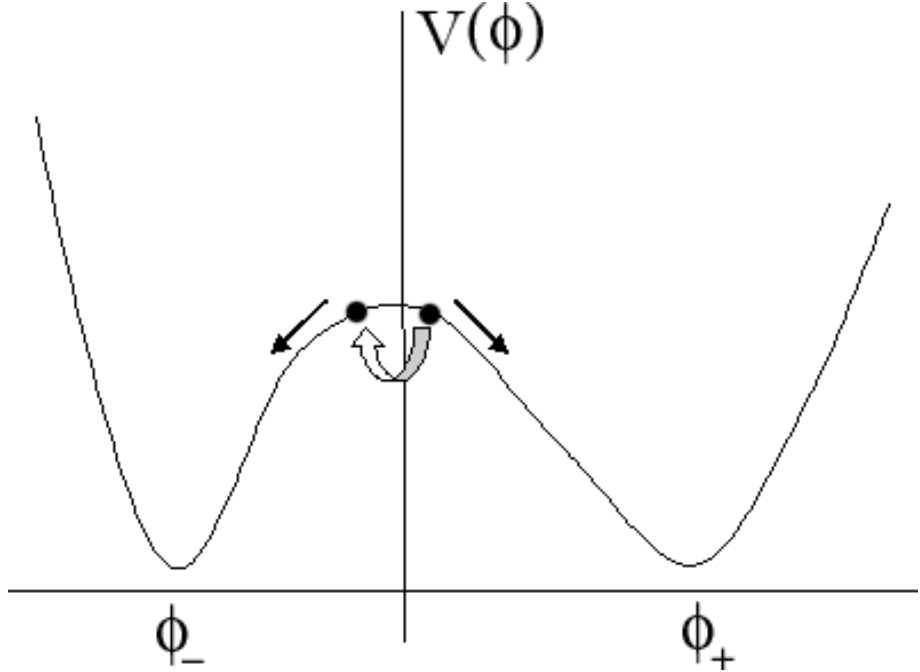}
\end{center}
\caption{Quantum creation of walls during inflation. Right black
point relates to the initial field value, $\phi _{in}$}
 \label{fluctwall}
\end{figure}

Let us choose a positive value for the initial field,
$\phi_{\rm{in}}>0$, as illustrated in the Fig.~\ref{fluctwall}. Then
an average field value will increase with time, ultimately reaching
a minimum of the potential at some value $\phi_{+}>0$. It means that
a greater part of space will be finally filled with the field value
$\phi=\phi_{+}$. Meanwhile, the field in some (small) space domain
could jump over the maximum due to the quantum fluctuations. In the
following, an average value of the field representing this
fluctuation tends to an another minimum of the potential,
$\phi_{-}<0$. As the result, the space at the final stage will be
filled by vacuum $\phi_{+}$ while some space domain is characterized
by the field value $\phi=\phi_{-}< 0$. If one starts to move from
inside of the domain to the outside, the path would start from a
space point with $\phi_{-}$ and finish at a space point with
$\phi_{+}$. Hence, the path must contain a point with the maximum
value of potential. It means that a wall is formed inevitably
between such space domains and the ``outer'' space with
$\phi=\phi_{+}$ \cite{Book1,PBH}.

\begin{figure}[ptb]
\begin{center}
\includegraphics[width=0.75\textwidth]{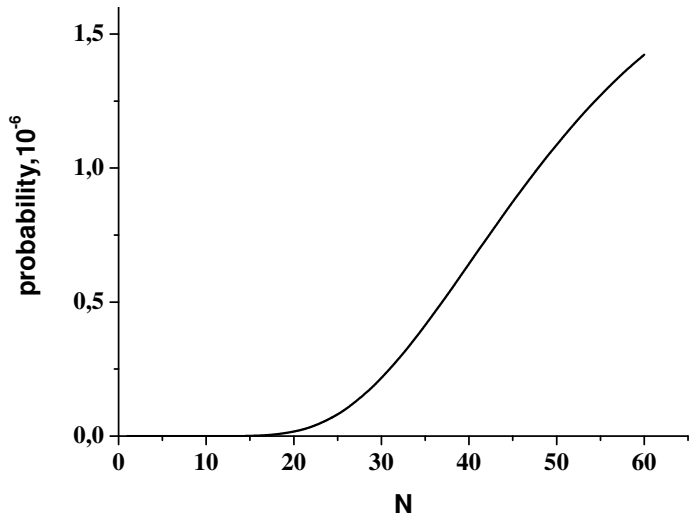}
\end{center} \caption{Part of space occupied by an another vacuum
state, depending on the time measured by e-folds. The initial field
value is $\phi_{\rm{in}} = \phi_{\max} + 3H$, where $\phi^{\max}$ is
a field value at a maximum of the potential. The parameter $m$ is
chosen to be $m=0.3H$, where $H$ is the Hubble parameter at the top
of potential.}
 \label{ratio}
\end{figure}

The ``dangerous'' values of fluctuations are those with $\phi\leq0$.
Such space domains will be surrounded by closed walls and if their
number is too large it would strongly influence the dynamics of the
early Universe. It is useful to calculate the probability of
nucleation of these fluctuations. The latter depends on the initial
field value $\phi_{\rm{in}}$ at the moment of creation of our
Universe. The corresponding probability
\begin{equation}
 P_{0}(\phi_{\rm{in}},T)=
 \int_{\phi=-\infty}^{\phi=0}\!dP(\phi,T;\phi_{\rm{in}},0)
\end{equation}
to find the field value $\phi\leq0$ at some space point for
reasonable values of parameters is represented in Fig.~\ref{ratio}.
This probability determines a ratio of spatial volumes with
different signs of the field. This probability is highly sensitive
to the initial field value $\phi_{\rm{in}}$: a closer to the
potential maximum it is nucleated, a greater part of the Universe
will be covered with walls at the final stage.

If a fraction of space surrounded by the walls is not very large,
the resulting massive BHs, which are formed from the walls, could
explain the early formation of quasars \cite{DER}. In opposite case
a contribution of BHs into the DM density is unacceptably large
\cite{PBH}. A further increasing of wall content leads to the
wall-dominated Universe \cite{Zeldovich74}.

A mass and space distribution of resulting BHs strongly depends on a
specific model and parameters of involved Lagrangian. It is
instructive to demostrate the mechanism of massive primordial BHs
production in the framework of the hybrid inflation model
\cite{LindeHyb} following to the results of \cite{PBH}.

\subsubsection{SBH formation in the Hybrid Inflation}

\begin{figure}
\begin{center}
\includegraphics[scale=0.25]{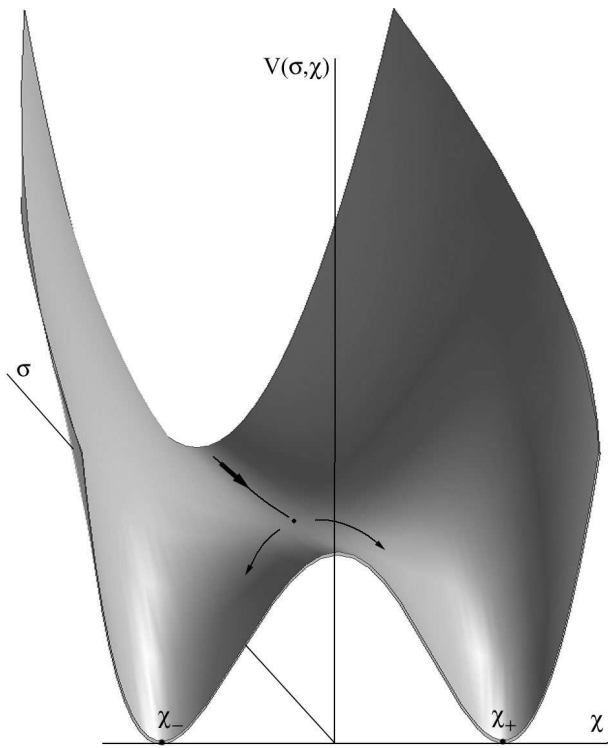} \end{center}
\caption{A form of potential for the hybrid inflation model. Arrows
indicate directions of classical motion of the fields. A black dot
indicates the critical point.}
 \label{Hybrid}
\end{figure}

A potential of the hybrid inflation model according to
\cite{LindeHyb} has the form
\begin{equation}
 V(\chi,\sigma)=
 \varkappa^{2}\left(M^{2}-\frac{\chi^{2}}{4}\right)^{2}+
 \frac{\lambda^{2}}{4}\chi^{2}\sigma^{2}+\frac{1}{2}m^{2}\sigma^{2}.
 \label{hpotential}
\end{equation}
Inflationary expansion of the Universe takes place during a slow
rolling along the valley $\chi =0,\; \sigma >\sigma _{c}$. When a
field $\sigma$ passes through the critical point
\[
\sigma _{c}=\sqrt{2}\frac{\varkappa }{\lambda }M,
\]
a motion along the line $\chi=0,\; \sigma<\sigma_{c}$ becomes
unstable. As a result, a field $\chi$ quickly moves to one of the
minima, $\chi_{\pm}=\pm2M$, $\sigma =0$, see Fig.~\ref{Hybrid} where
potential (\ref{hpotential}) for the hybrid inflation model is
represented. Inflation is finished with the intensive field
fluctuations around one of an accidentally chosen minimum.

It is clear now that this well elaborated picture suffers a serious
problem nevertheless. During an inflationary stage, when the fields
$\sigma$ and $\chi$ move classically along the line $\chi =0$, the
space is divided into many causally disconnected regions. The values
of scalar fields within these regions are slightly different due to
quantum fluctuations. An each e-fold produces approximately
$e^{3}\simeq 20$ space domains. Hence there are about
$e^{180}\simeq10^{78}$ space domains right before the end of
inflation. The values of field are chaotically distributed around
the point $\chi =0,\; \sigma=\sigma_{c}$. The domains with a field
value $\chi <0$ tend to the left minimum $\chi_{-}=-2M,\; \sigma=0$.
Another part fall to the right minimum $\chi _{+}=+ 2M,\; \sigma=0$.
A lot of walls between these domains appear and we come to a well
known problem of the wall-dominated Universe \cite{Zeldovich74}.

The only way for our Universe to evolve into the modern state is to
be created with an initial field value $\chi_{\rm in}\neq 0$ at the
beginning of inflation. During inflation, the field $\chi$ must
slowly approach to the critical line $\chi =0$. If, in the middle of
inflation, an average field value approaches to the critical line
$\chi =0$, the field in some part of space domains could cross this
line due to fluctuations. In the future, these domains will be
filled by vacuum, say, $\chi_{-}$ surrounded by a sea of an another
vacuum $\chi_{+}$. These two vacua are separated by a closed wall as
it was discussed above. A number of these walls strictly depends on
the initial conditions at the creation moment of our Universe.

Let us estimate the energy and size of closed walls in the framework
of the hybrid inflation. To this end, let us suppose that a field in
some volume crossed the critical line at e-folds number $N$ before
the end of inflation. Its size is about the Hubble radius, $H^{-1}$,
and it will be increased in $e^{N}$ times during inflation. A
surface energy density of the domain wall after inflation defined by
the potential (\ref{hpotential}) is
\begin{equation}
 \epsilon =\frac{8\sqrt{2}}{3}\varkappa M^{3}.
  \label{sigma}
\end{equation}
Thus an energy $E_{\rm wall}$ contained in the wall after inflation
is at least
\begin{equation}
E_{\rm wall}\simeq 4\pi \epsilon \left( H^{-1}e^{N}\right)
^{2}=4\sqrt{2}\frac{ M_{\rm Pl}^{2}}{\varkappa M}e^{2N}.
 \label{Ewall}
\end{equation}
A numerical value $N$\ varies in the interval $(0<N<N_{U}\simeq
60)$. These walls will collapse into the BHs with mass $M_{\rm
BH}\simeq E_{\rm wall}$ (for details see \cite{Ru01b}).

Let us estimate masses of these BHs for a representative values of
parameters $\varkappa=10^{-2}$ and $M=10^{16}$~GeV. If $N=40$, we
obtain for the mass of formed BH
$$
M_{\rm BH}\simeq 3\, 10^{59}\mbox{~GeV}\sim 100{\rm M}_{\odot}.
$$
A similar estimation for a smallest BHs, which are created at the
e-fold number $N=1$ before the end of inflation, gives
$$
M_{\rm BH,small}\simeq 10^{6}M_{\rm Pl}.
$$
Therefore, the hybrid inflation leads to the formation of BHs with
mass in a rather wide range, $M_{\rm BH}>10^{25}$~GeV with an upper
limit of $\sim10^2{\rm M_{\odot}}$. An amount of massive BHs depends
on how close an average field value approaches to the critical line
$\chi =0$. The last, in turn, depends on the initial conditions and
specific values of parameters of this model.

\subsubsection{Evolution of massive cluster of primordial BHs}

The main lesson of previous discussion is that the massive
primordial BHs are formed after inflation as a rule rather than as
an exception. As was shown in \cite{Ru05} the mechanism described
above leads to the nucleation of groups (clusters) of primordial BHs
rather than to a single primordial BH. Here we discuss the idea that
these clusters of primordial BHs could be the seeds for early galaxy
formation \cite{DER}. Our consideration is based on the results
published in \cite{Ru1,Ru01b,Book1}, where the ``Mexican hat''
potential for the scalar field was considered. The calculations with
the ``Mexican hat'' potential give the plausible parameters of
clusters of primordial BHs.

In numerical calculations, the cluster of primordial BHs is modelled
by a spherically symmetric system with a radius $r<ct$, consisting
of primordial BHs with a total mass $M_h$ inside the radius $r$. To
fix the cosmological evolution after inflation it is also supposed
that the density of radiation is $\rho_r$, the density of ordinary
DM is $\rho_{\rm{DM}}$ and the density of $\Lambda$-term is
$\rho_{\Lambda}$. The main, most massive BH in a chosen cluster is
in the center of the system. The density of radiation (and obviously
the density of $\Lambda$-term) is homogenous. Therefore, the
fluctuations induced by the primordial BHs are classified as entropy
fluctuations. The scale under consideration is smaller than the
horizon scale, and we use the Newtonian gravity with the
prescription \cite{TolMc30} of treating the gravitation of
homogenous relativistic components, $\rho\to\rho+3p/c^2$.

The evolution of any fictitious spherical shell of considered
cluster with an initial radius $r_i <r$ obeys the equation
\begin{equation}
 \frac{d^2r}{dt^2}=
 -\frac{G(M_h+M_{\rm{DM}})}{r^2}-\frac{8\pi G\rho_r r}{3}+
 \frac{8\pi G\rho_{\Lambda} r}{3},
 \label{d2rdt1}
\end{equation}
with an approximate initial conditions at the moment $t_i$: $\dot
r=Hr$, $r(t_i)=r_i$. In obtaining equation (\ref{d2rdt1}), it was
taken into account that $\varepsilon_r+3p_r=2\varepsilon_r$,
$\varepsilon_{\Lambda}+3p_{\Lambda}=-2\varepsilon_{\Lambda}$.
Numerical calculations on the basis of equation (\ref{d2rdt1}) with
an initial conditions indicate that the radius of the shell
increases initially, and gradually its expansion decelerates. At
some instant, a speed of expansion diminishes to zero. The shell is
separated from the cosmological expansion and starts to shrink. All
components of nonrelativistic matter --- the DM, primordial BHs and
baryons follow the dynamics of the shell surrounding them. As a
result, the solutions of (\ref{d2rdt1}) for shells with different
initial radii provide us the density distribution of the DM and
primordial BHs in the evolving cluster.

It is suitable to rewrite the main equation by using a new variable
$b(t)$.
\begin{equation}
 \label{bt}
 r(t)=\xi a(t)b(t)
\end{equation}
In this parametrization, $\xi$ is a comoving length, $a(t)$ is a
scale factor of the Universe normalized to the present moment $t_0$
as $a(t_0)=1$ and the function $b(t)$ characterizes the deflection
of the cosmological expansion from the Hubble law. The variable
$\xi$ is connected to the mass of DM inside the considered spherical
volume (excluding total BHs mass) by the relation
$M_{\rm{DM}}=(4\pi/3)\rho_{\rm{DM}}(t_0)\xi^3$, where $\rho_{\rm
DM}(t_0)$ is the modern DM density. The function $a(t)$ obeys the
Friedman equation, which can be rewritten as $\dot a/a=H_0E(z)$,
where $z=a^{-1}-1$ is the redshift, $H_0$ is the current value of
the Hubble constant and function $E(z)$ has a form
\begin{equation}
 E(z)=[\Omega_{r,0}(1+z)^4+\Omega_{m,0}(1+z)^3+
 \Omega_{\Lambda,0}]^{1/2},
 \label{efun}
\end{equation}
where $\Omega_{r,0}$ is a density parameter of radiation,
$\Omega_{m,0}\simeq0.3$, $\Omega_{\Lambda,0}\simeq0.7$, and $h=0.7$.
Using the second Friedman equation (for $\ddot a$), one can rewrite
(\ref{d2rdt1}) as follows
\begin{equation}
 \frac{d^2b}{dz^2}+\frac{db}{dz}S(z)+
 \left(\frac{1+\delta_h}{b^2}-b\right)
 \frac{\Omega_{m,0}(1+z)}{2E^2(z)}=0,
 \label{d2bdz1}
\end{equation}
where function
\begin{equation}
S(z)=\frac{1}{E(z)}\frac{dE(z)}{dz}-\frac{1}{1+z}
\end{equation}
and a value of fluctuation $\delta_h=M_h/M_{\rm{DM}}$. Equation
(\ref{d2bdz1}) is equivalent to those obtained in \cite{kt} in the
case $\Omega_{\Lambda}=0$. We start tracing the evolution of cluster
at a high redshift $z_i$ when the considered shell crosses the
horizon $r\sim ct$.

\begin{figure}[ptb]
\begin{center}
\includegraphics[angle=0,width=0.95\textwidth]{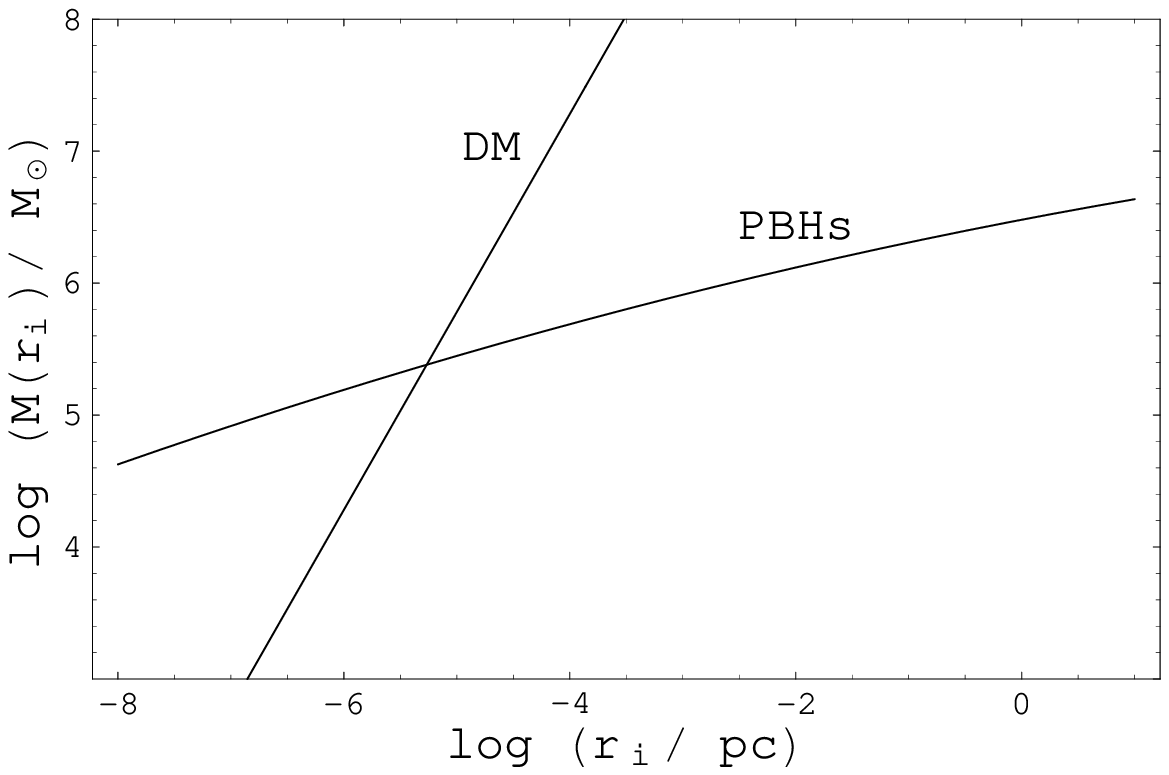} \\
\includegraphics[angle=0,width=0.95\textwidth]{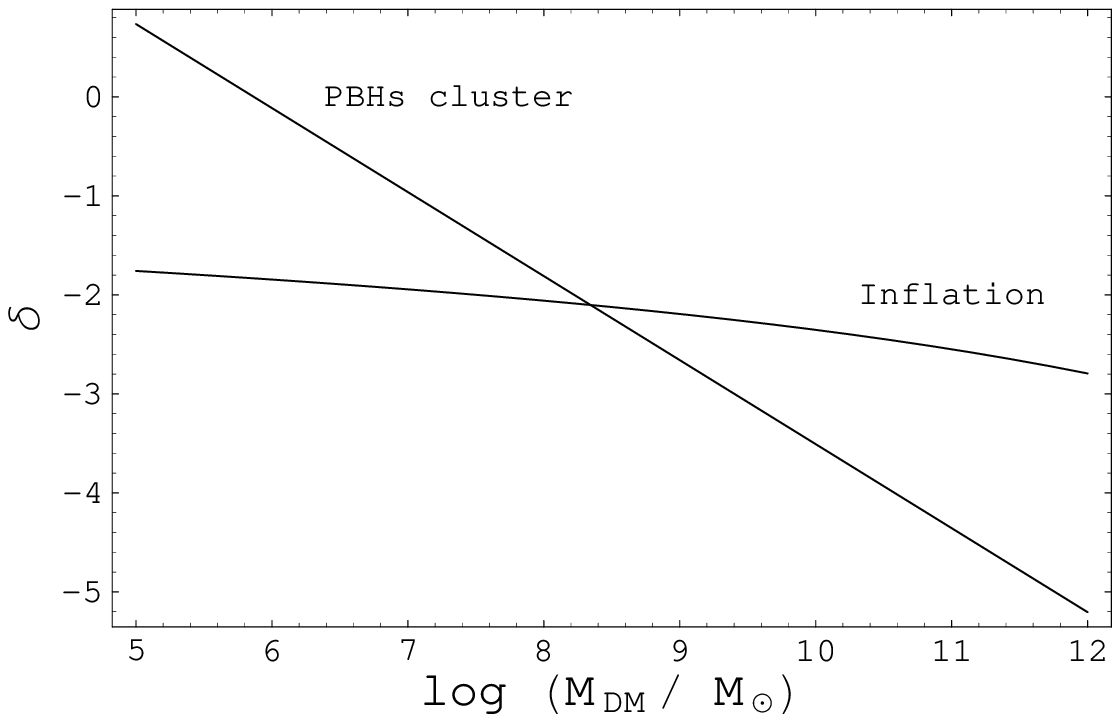}
\end{center}
\caption{In the top panel the initial mass profile $M_{h}(r_{i})$ of
the cluster of primordial BHs and the mass profile
$M_{\rm{DM}}(r_{i})$ of DM are shown. In the bottom panel, the
r.m.s. density perturbation at the moment $t_{\rm{eq}}$ of
matter-radiation equality is shown. Two cases are compared: the
total density fluctuations produced in the presence of the cluster
of primordial BHs and those produced in a standard way as remnants
of the inflationary stage.}
 \label{massprof}
\end{figure}

Initial mass profile $M_{h}(r_{i})$ of the cluster of primordial BHs
was calculated following to formalism developed in \cite{Ru05} and
has the form presented in the Fig.~\ref{massprof} (upper panel).
This numerical result is a starting point for the following
analysis. The mass $M_{\rm{DM}}(r_{i})$ of DM inside the same
spherical shells is also shown for a comparison. The radius $r_{i}$
is the physical size of sphere at the moment $t_{i}$ and the
temperature $T_{i}$ corresponds to the moment when the sphere is
crossing the cosmological horizon. Note that the radiuses of shells
in Fig.~\ref{massprof} are taken at different instants $t_{i}$.
Therefore, the shown mass of uniformly distributed DM not follows to
the law $M_{\rm{DM}}\propto r^{3}$ as it must be for a fixed moment,
common for all spheres. Physical size at the temperature $T_{i}$ is
smaller than that in the modern epoch in $T_{0}/T_{i}$ times, where
$T_{0}=2.7$~K. A density in the local center is so high that a lot
of primordial BHs appear to be inside their common gravitational
radius $r_g=2GM/c^2$. According to numerical calculations with
specific values of parameters, a total mass of primordial BHs
appears inside the horizon from the beginning is $2.7\,10^4{\rm
M}_{\odot}$. Hence, the massive BH with this mass is formed at
first.

A widespread prejudice is that a mass of formed primordial BH could
not exceed a total mass under the horizon at the same time, $M\sim
(t/t_{\rm Pl})M_{\rm Pl}$ \cite{Carr94}. As a result, the mass of
primordial BH seems must be not larger than ~$10^3 {\rm M}_{\odot}$.
It is true indeed if one considers the formation of primordial BHs
density fluctuations. On the contrary, in our case the mass of the
formed primordial BHs is determined by the area of the closed walls,
formed and stretched during inflation. Size of wall could be greater
its horizon size in orders of magnitude. BH is nucleated when the
wall collapse under its horizon.

The following dynamics of the cluster of primordial BHs together
with DM component could finally lead to the formation of galaxies
provided that the numerous collisions of galaxies are taken into
account. Results of calculations made below are applicable both for
an inner part of the cluster, composed mainly from primordial BHs
and collapsing at the radiation dominated stage, and for an outer
regions of the cluster, where DM is a main dynamical component. The
outer regions are detached from the cosmological expansion at the
matter dominated epoch.

The most early epoch in our calculations corresponds to the
formation of central BH with a mass $2.7\,10^4{\rm M}_{\odot}$
described above. The temperature of the Universe at that time is
$\simeq16$~MeV. We suppose that DM has been already decoupled from
radiation at this temperature. For example, neutralino DM particles
with mass $100$~GeV and slepton with mass $1$~TeV decouples at the
temperature $\simeq150$~MeV \cite{Schw03}, well before the
considered epoch. Therefore, the neutralino DM can be treated as
moving freely under the influence of gravitational force only. As a
result the clustering of two-component medium (BHs$+$DM) is
described by the equation (\ref{d2bdz1}) from the very beginning.
The same situation is realized for DM composed of the super-heavy
particles with mass $m_{\chi}\sim10^{13}-10^{14}$~GeV, which
probably never been in the kinetic equilibrium with radiation. In
the opposite case (for some another DM particles candidates), the
growth of fluctuations in DM medium is suppressed by the friction on
radiation until the kinetic decoupling, while the BHs are
clustering. The super-heavy particles are preferable for our model
in comparison with the neutralino because their annihilation
cross-section is very small, $\propto m_{\chi}^{-2}$. In this case
the problems with a huge annihilation signal from the considered
protogalaxies are absent.

All scenarios imply a further merging of BHs during the multiple
coalescences of host minihalos
\cite{merging,merging2,merging3,merging4} accompanied also by an
additional gas accretion. It is worth to estimate the probability to
find a nowadays galaxy without the SBH in the framework of the
scenario discussed above. Induced galaxies containing massive BHs
and ordinary small protogalaxies without BHs have masses around
$M_{\rm{DM}}=10^{8}{\rm M}_{\odot}$, while nowadays galaxies are as
massive as $M_{\rm{DM}}=10^{12}{\rm M}_{\odot}$. Every collision of
an induced galaxy with an ordinary protogalaxy produces the next
generation of protogalaxy with a massive BH in its center. So about
$10^{4}$ collisions have happened up to now. Suppose that the amount
of induced galaxies is about 0.1\% in comparison with the ordinary
ones. Upper limit of probability to find a nowadays galaxy without a
SBH is $0.999^{10^{4}}\simeq 4.5\,10^{-5}$. Hence, even a very small
amount of induced galaxies is able to explain the observable
abundance of e.~g. the AGN.

\section{Conclusions and Discussions}

The existence of SBHs at the centers of galaxies looks like an
unavoidable fact. The SBHs play an important role in the formation
and global evolution of galaxies and the intergalactic medium.
Nevertheless a valuable theory of SBHs formation is still absent. It
appears not so easy to explain the whole set of observations which
are biased and sometimes contradictive. The SBHs mass correlate with
the mass of the host galaxies, though exceptions are known. For
example, a suspected SBH was found recently in the dwarf galaxy
VCC128 \cite{Deb}. The maximum accretion rate onto the SBHs, at
redshift $z\sim 0.7$ almost coincides with a peak density of
luminous infrared galaxies at $z\sim 0.8$. One could expect a quick
growth of already existing SBHs. At the same time, the most massive
BHs are found much earlier, at $z>6.0$.

At the recent epoch we are witnesses of impressive competition of
two main directions in BH investigations. Several decades ago an
idea was accepted that all BHs, if exist, result from the supernova
explosions. Formation of galaxies seemed definitely precede a
formation of BHs. But gradually this picture becomes more and more
tangled. New observational data during the last 20 years indicate
that almost all massive galaxies contain SBHs with a mass in the
range $10^6-10^8{\rm M}_{\odot}$. Moreover, the discovery of quasars
at high redshifts means a combined evolution of the galactic nuclei
and SBHs. The ubiquity of SBHs in galaxies may be the indication of
some special and universal mechanism for SBH formation:
simultaneously with the galaxies or even before the formation of
galaxies.

The nowadays observations reveal a very early formation of SBHs:
they already formed at the epoch of 800 million years after the Big
Bang. Considerable efforts are needed to explain so early existence
of SBHs originated from the remnants of stars. It seems that an idea
of Population III stars with mass exceeding $10^2{\rm M}_{\odot}$ is
able to solve the problem. Nevertheless, it is not excluded that
astrophysical scenario for SBH formation will not succeed.

Meantime new direction of primordial BHs research becomes more and
more popular. The idea of primordial BHs is not very new --- the
first paper were published in 1966, \cite{zeldnov66,hawking71}. It
was supposed that primordial BHs were produced from density
fluctuations long before the star formation. These BHs are very
light, $M_{\rm BH}\ll 10^{15}$~g, and should be quickly evaporated
due to the Hawking radiation. These BHs could not merge
effectively into the rather massive ones. Nevertheless, an idea of
primordial BHs was survived. Several absolutely different
mechanisms of massive and even supermassive primordial BHs
formation has been revealed and widely discussed
\cite{Carr,Star92,ivan94,Jedamzik,Khlopov80,Bubble,Bubble2,Bubble3,%
Bubble4,Bubble5,Bubble6,Bubble7,Yoko,Ru25,Ru1,Ru01b,Ru05}.

Competition between astrophysical and cosmological schemes of SBHs
origin is going on. New observational data are analyzed from the
both points of view. Both astrophysical and cosmological scenarios
posses as advantages and some defects. The astrophysical scenarios
of SBHs formation has rather firm basis but suffer from the lack of
time and shortage of matter supply for building the SBHs in galaxies
by means of accretion and merging. Meanwhile, the cosmological
scenarios are based only on the supposed mechanisms but provide the
way of early SBHs formation. It is not excluded that both scenarios
of SBHs formation were realized in the Universe. More elaborate and
predictive models and new detailed observations are requested to
resolve the SBH enigma. The upcoming gravitational wave laser
interferometers provide a very promising new channel for the
exploration of SBH problem.

Acknowledgements: This work was supported in part by the Russian
Foundation for Basic Research grants 06-02-16029 and 06-02-16342,
and the Russian President grants LSS 4407.2006.2 and LSS
5573.2006.2.


\begin{thebibliography}{99}

\bibitem{begblrees84} Begelman, M.C., Blandford, R.D., and Rees,
M.C., 1984, Rev. Mod. Phys., 56, 255.

\bibitem{Rees84} Rees, M.J. 1984, Ann. Rev. Astron. Astrophys.,
22, 471.

\bibitem{Antonucci} Antonucci, R.R.J. 1993, Ann. Rev. Astron.
Astrophys., 31, 473.

\bibitem{FerrareseFord04} Ferrarese, L., and Ford, H. 2004,
{\tt arXiv:astro-ph/0411247}.

\bibitem{galacticbh} Falcke, H., and Hehl, F.W. (Eds.) 2003, The
Galactic black holes, Series in high energy physics, cosmology and
gravitation. IOP Publishing Ltd, London.

\bibitem {Gnedin01} Gnedin, O.Y. 2001, Class. Quant. Grav., 18,
3983.

\bibitem{Li} Li, Y., et al. 2006, {\tt arXiv:astro-ph/0608190}.

\bibitem{Fan} Fan, X., et al. 2003, Astron. J., 125, 1649.

\bibitem{Schw} K. Schwarzschild, 1916, Sitzungsber. Preuss. Akad.
Wiss., Berlin (Math. Phys.) 189; (English translation in {\tt
arXiv:physics/9905030}).

\bibitem{petersmat63} Peters, P.S., and Mathews, J. 1963, Phys.
Rev. D, 131, 435.

\bibitem{Mag05} Magain, P. et al. 2005, Nature 437, 381.

\bibitem{HaeDavRee05} Haehnelt, M.G., Davies, M.B., and Rees,
M.J. 2006, Mon. Not. Roy. Astron. Soc., 366, L22.

\bibitem{CavVit01} Cavaliere, A., and Vittorini, V. 2001,
{\tt arXiv:astro-ph/0110644}.

\bibitem{haehreeesnar98}  Haehnelt, M.G., Natarajan, P., and Rees,
M.J. 1998, Mon. Not. Roy. Astron. Soc., 300, 817.

\bibitem{KauHae00} Kauffmann, G., and Haehnelt, M.G. 2000,
Mon. Not. Roy. Astron. Soc., 311, 576.

\bibitem{choi00} Choi, Y., Yang, J., and Yi, I. 2000,
{\tt arXiv:astro-ph/0005590}.

\bibitem{hopkins05} Hopkins, P.F. et al. 2005, Astrophys. J., 630,
705.

\bibitem{seed06} Mack, K.J., Ostriker, J.P., and Ricotti, M. 2006,
{\tt arXiv:astro-ph/0608642}.

\bibitem{MDS06} Marziani, P., Dultzin-Hacyan, D., Jack, W., and
Sulentic, J.W. 2006, in New Development in Black Hole Research,
Kreitler P.V. (Editor), New York: Nova Science Publishers, pp.
123-183.

\bibitem{Tal05} Tal, A. 2006, Phys. Rep., 419, 65.

\bibitem{Gez06}Gezari, et al. 2006, Astrophys. J., 653, L25.

\bibitem{Rice} Rice, W.K.M., Lodato, G., and Armitage, P.J. 2005,
Mon. Not. Roy. Astron. Soc., 364, L56.

\bibitem{Paumard} Paumard, T. et al. 2006, Astrophys. J., 643,
1011.

\bibitem{Nayakshin} Nayakshin, S. 2007, {\tt
arXiv:astro-ph/0701150}.

\bibitem{Weaver} Weaver, H., et al. 1965, Nature, 208, 29.

\bibitem{Churchwell} Churchwell, E., et al. 1977, Astron.
Astrophys., 54, 969.

\bibitem{Kondratko} Kondratko, P.T., Greenhill, L.J., and Moran
2006, Astrophys. J., 652, 136.

\bibitem{Nandra} Nandra, K., et al. 1997, Astrophys. J., 477,
602.

\bibitem{chenhalp} Chen, K., and Halpern, J.P. 1998,
Astrophys. J., 344, 115.

\bibitem{fabreessw} Fabian, A.C., Rees, M., Stella, L., and White, N.E.
1989, Mon. Not. Roy. Astron. Soc., 238, 729.

\bibitem{matt} Matt, G., Perola, G.C., and Stella, L. 1993,
Astron. Astrophys., 267, 643.

\bibitem{zakharov1} Zakharov, A.F. 1991, Soviet Astron., 35, 30.

\bibitem{fabiwary} Fabian, A.C., Iwazawa, K., Reynolds, C.S., and
Young, A.J. 2000, Publ. Astron. Soc. Pac., 112, 1145.

\bibitem{zakharov2} Zakharov, A.F. et al. 2005, {\tt
arXiv:gr-qc/0507118}.

\bibitem{pariev1} Pariev, V.I., and Bromley, B.C. 1998,
Astrophys. J., 508, 590.

\bibitem{pariev2} Pariev, V.I., Bromley, B.C., and Miller, W.A.
2001, Astrophys. J., 547, 649.

\bibitem {Shen} Shen, Z.-Q., Lo, K.Y., Liang, M.-C., Ho, P.T.P.,
and Zhao, J.H. 2005, Nature, 438, 62.

\bibitem{KorRich05} Kormendy, J., and Richstone, D.O. 1995, Ann.
Rev. Astron. Astrophys., 33, 581.

\bibitem{mag98} Magorrian, J. et al. 1998, Astron. J, 115, 2285.

\bibitem{kazant05} Kazantzidis, S. et al. 2005, Astrophys. J.,
623, L67.

\bibitem{Gebh} Gebhardt, K. et al. 2000, Asrtophys. J., 539, L642.

\bibitem{Ferra} Ferrarese, L., and Merritt, D. 2000, Astrophys. J.,
539, L9.

\bibitem{Tre02} Tremaine, S. et al. 2002, Astrophys. J., 574, 740.

\bibitem{SSGB06} Salviander, S., Shields, G.A., Gebhardt,  K., and
Bonning, E.W. 2007, Astrophys. J., 662, 131.

\bibitem{MerDunForTer02} Merrifield, M.R., Forbes, D.A., and
Terlevich, A.I. 2000, Mon. Not. Roy. Astron. Soc., 313, L29.

\bibitem{SilRee98} Silk, J., and Rees, M.J. 1998, Astron.
Astrophys., 331, L4.

\bibitem{HaeKau001} Haehnelt, M.G., and Kauffmann, G. 2000,
Mon. Not. Roy. Astron. Soc., 318, L35.

\bibitem{DokEro03} Dokuchaev, V.I., and Eroshenko, Yu.N. 2003,
Astron. Astrophys. Trans., 22, 727.

\bibitem{FukTur96} Fukugita, M., and Turner, E.L.
1996, Astrophys. J., 460, L81.

\bibitem{Kaa01} Kaaret, P. et al. 2001, Mon. Not. Roy. Astron.
Soc.,  321, L29.

\bibitem{MouTan02} Mouri, H., and Taniguchi, Y. 2002, Astrophys.
J., 566, L17.

\bibitem{GebRicHo02} Gebhardt, K., Rich R.M., and Ho L.C. 2002,
Astrophys. J.,  578, L41.

\bibitem{Tre06} Trenti, M. 2006, {\tt arXiv:astro-ph/0612040}.

\bibitem{vdM02} van der Marel, R.P. et al. 2002, {\tt
arXiv:astro-ph/0209314}.

\bibitem{dokero2002} Dokuchaev, V.I., and Eroshenko, Yu.N. 2002,
Zh. Eksp. Teor Fiz, 121, 5 (JETP, 2002 94, 1).

\bibitem{ufn2} Gurevich, A.V., Zybin, K. P., and Sirota, V.A.
1997, Uspekhi Fiz, Nauk, 40, 913 (Physics --- Uspekhi, 1997, 40,
869).

\bibitem{Haehnelt} Haehnelt, M.G. 1994, Mon. Not. Roy. Astron.
Soc., 269, 199; {\sc ibid.} 2003, Class. Quant. Grav., 20, S31.

\bibitem{MenHaiNar01} Menou, K., Haiman, Z., and Narayanan, V.K.
2001, Astrophys. J., 558, 535.

\bibitem{WyiLoe02} Wyithe, J.S.B., and Loeb, A. 2003, Astrophys.
J., 590, 691.

\bibitem{berti05} Berti, E., Buonanno, A., and Will, C.M. 2005,
Class. Quant. Grav., 22, S943.

\bibitem{Moore93} Moore, B. 1993, Astrophys. J. Lett., 413,  93.

\bibitem{carr75} Carr, B.J. 1975, Astrophys. J., 201, 1.

\bibitem{nemir} Nemiroff, R.J., Marani, G.F., Norris, J.P., and
Bonnel, J.T. 2001, Phys. Rev. Lett., 86, 580.

\bibitem{wilk} Wilkinson, P.N. et al. 2001, Phys.  Rev. Lett., 86,
584.

\bibitem{GonSil99} Gondolo, P., and Silk, J. 1999, Phys. Rev.
Lett., 83, 1719.

\bibitem{UliZhaKam01} Ulio, P., Zhao, H., and Kamionkowski, M.
2001, Phys. Rev. D, 64, 043504.

\bibitem{Mer02} Merritt, D., Milosavljevic, M, Verde, L., and
Jimenez, R. 2002, Phys. Rev. Lett., 88, 191301.

\bibitem{ZhaSil05} Zhao, H., and Silk, J. 2005, Phys. Rev. Lett.,
95, 011301.

\bibitem{MacHen03} MacMillan, J.D., and Henriksen, R.N.
2002, Astrophys. J., 569, 83.

\bibitem{spitzer} Spitzer, L. Jr. 1987, Dynamical evolution of
Globular Clusters, Princeton Univ. Press, Princeton, New Jersy.

\bibitem{saslaw} Saslaw, W. 1987, Gravtational physics of stellar
and galactic systems, Cambridge Univ. Press, Cambridge.

\bibitem {dokrev} Dokuchaev, V.I. 1991. Soviet Phys. Usp., 34, 447
(1991, Usp. Fiz. Nauk, 161, 1).

\bibitem{ss66} Spitzer, L. Jr., and Saslaw W. 1966, Astrophys. J.,
143, 400.

\bibitem{sanders70} Sanders, R.H. 1970, Astrophys. J., 162, 791.

\bibitem{spitzerh71} Spitzer, L., and Hart, M.H. 1971, Astrophys.
J., 164, 399.

\bibitem{MerRep06} Merritt, D. 2006, Rep. Prog. Phys., D69, 2513.

\bibitem{fpr75} Fabian, A.C., Pringle, J.E., and Rees, M.J. 1975,
Mon. Not. Roy. Astron. Soc., 172, 15P.

\bibitem{heggie75} Heggie, D.C. 1975, Mon. Not. Roy. Astron. Soc.,
173, 729.

\bibitem{pressteuc77} Press, W.H., and Teulolsky, S.A. 1977,
Astrophys. J., 213, 183.

\bibitem{begrees78} Begelman, M.C., and Rees, M.J. 1978,
Mon. Not. Roy. Astron. Soc., 185, 847.

\bibitem{QuiSha87} Quinlan, G.D., and Shapiro, S.L. 1987,
Astrophys. J., 321, 199.

\bibitem{QuiSha89} Quinlan, G.D., and Shapiro, S.L. 1989,
Astrophys. J., 343, 725.

\bibitem{QuiSha90} Quinlan, G.D., and Shapiro, S.L. 1990,
Astrophys. J., 356, 483.

\bibitem{berdok01} Berezinsky, V.S.,  and Dokuchaev, V.I. 2001,
Astropart. Phys., 15, 87.

\bibitem{berdok06} Berezinsky, V.S., and Dokuchaev,  V.I. 2006,
Astron. Ap., 454, 401.

\bibitem{llmech} Landau, L.D., and Lifshitz, E.M. 1960, Mechanics,
Eddison-Wesley, Reading Mass.

\bibitem{chandra43} Chandrasekhar, S. 1943, Rev. Mod. Phys., 15,
21.

\bibitem{spitharm58} Spitzer, L., and Harm, R. 1958, Astrophys. J.,
127, 544.

\bibitem{lightshap78} Lightman, A.P., and Shapiro S.L. 1978, Rev.
Mod. Phys., 50, 437.

\bibitem{milgshap78} Milgrom, M., and Shapiro S.L. 1978,
Astrophys. J., 223, 991.

\bibitem{ozdok82} Ozernoy, L.M., and Dokuchaev, V.I. 1982,
Astron. Ap., 111, 1.

\bibitem{dokoz82} Dokuchaev, V.I., and Ozernoy, L.M. 1982, Astron.
Ap., 111, 16.

\bibitem{hoylefowler63} Hoyle, F., and Fowler, W. 1963, Nature,
197, 533.

\bibitem{fowler66} Fowler, W. 1966, Astrophys. J., 144, 180.

\bibitem{fowler66b} Fowler, W. 1966, Rev. Mod. Phys., 36, 545
(erratum 36, 1104).

\bibitem{zeldnov66} Zel'dovich, Ya.B., and Novikov, I.D. 1966,
Astron. Zh., 43, 758 (Sov. Astron. --- A.~J., 1976, 10, 602).

\bibitem{hawking71} Hawking, S. W. 1971, Mon. Not. Roy. Astron.
Soc., 152, 75.

\bibitem{zeldnov71} Zel'dovich, Ya.B., and Novikov, I.D. 1971,
Relativistic Astrophysics, Vol. 1: Stars and Relativity, Univ. of
Chicago Press, Chicago.

\bibitem{haehrees93} Haehnelt, M.G., and Rees, M.J. 1993,
Mon. Not. Roy. Astron. Soc., 263, 168.

\bibitem{eisloeb95} Eisenstein, D.J., and Loeb, A. 1995,
Astrophys. J., 443, 11.

\bibitem{iben63} Iben, I. 1963, Astrophys. J., 138, 1090.

\bibitem{chandra64} Chandrasekhar, S. 1964, Phys. Rev. Lett., 12,
114 (erratum 12, 437).

\bibitem{chandra64b} Chandrasekhar, S. 1964, Astrophys. J., 140,
417.

\bibitem{chandra65} Chandrasekhar, S. 1965, Astrophys. J., 142,
1488.

\bibitem{lee87} Lee, H.M. 1987,  Astrophys. J., 319, 851.

\bibitem{sbss02} Saijo, M.T., Baumgarte, W., Shapiro, S.L., and
Shibata, M. 2002, Astrophys. J., 569, 349.

\bibitem{shibshapbss02} Shibata, M., and Shapiro, S.L. 2002,
Astrophys. J., 572, L39.

\bibitem{shibata03} Shibata, M. 2003, Astrophys. J., 595, 992.

\bibitem{ben92} Benz, W., and Hills, J.G. 1992, Astrophys. J., 389,
546.

\bibitem{lai93} Lai, D., Rasio, F.A., and Shapiro, S.L. 1993,
Astrophys. J., 412, 593.

\bibitem{shapteuk83} Shapiro, S.L., and Teukolsky, S.A.
1983, Black Holes, White Dwarfs and Neutron Stars, Willey, New-York.

\bibitem{new01} New, K.C., and Shapiro, S.L. 2001, Astrophys. J.,
548, 439.

\bibitem{kimbshap01} Kimberly, C.B., and  Shapiro, S.L. 2001,
Class. Quant. Grav., 18, 3965.

\bibitem{ZelPod65} Zel'dovich, Ya.B., and Podurets, M.A. 1965,
Astron. Zh., 42, 963 (Sov. Astron. --- A.~J., 1966, 9, 742).

\bibitem{ShaTeu} Shapiro, S.L., and Teukolsky, S.A. 1985,
Astrophys. J., 292, L41.

\bibitem{ShaTeu2} Shapiro, S.L., and Teukolsky, S.A. 1985,
Astrophys. J., 298, 34.

\bibitem{ShaTeu3} Shapiro, S.L., and Teukolsky, S.A. 1985,
Astrophys. J., 298, 58.

\bibitem{ShaTeu4} Shapiro, S.L., and Teukolsky, S.A. 1986,
Astrophys. J., 307, 575.

\bibitem {dokoz77} Dokuchaev, V.I., and Ozernoy, L.M. 1977, Pisma
Astron. Zh., 3, 391 (Sov. Astron. Lett., 1977, 3, 209).

\bibitem {dokrev2} Dokuchaev, V.I. 1991., Mon. Not. Roy. Astron.
Soc., 251, 564.

\bibitem{ilyin04} Ilyin, A.S., Zybin, K.P., and Gurevich, A.V. 2004,
Zh. Eksp. Teor. Fiz., 98, 5 (JETP, 2004, 98, 1).

\bibitem{benson05} Benson, A.J., Kamionkowski, M., and Hassani,
S.H. 2005, Mon. Not. Roy. Astron. Soc., 357, 847.

\bibitem{erickcek06} Erickcek, A.L., Kamionkowski, M., and Benson,
A.J. 2006, Mon. Not. Roy. Astron. Soc., 371, 1992.

\bibitem{TylJanSan03} Tyler, C., Janus, B., and Santos-Noble, D.
2003, {\tt arXiv:astro-ph/0309008}.

\bibitem{Rees92} Rees, M.J. 1992, in Physics of Active Galactic
Nuclei, eds. W.J. Duschl, S.J.~Wagner, Springer-Verlag, Berlin,
662.

\bibitem{Kawaguchi} Kawaguchi, T., Aoki, K., Ohta, K. and Collin,
S. 2004, Astron. Astrophys., 420, L23.

\bibitem{bde03} Babichev, E., Dokuchaev, V., and Eroshenko, Yu.
2004, Phys. Rev. Lett., 93, 021102.

\bibitem{bde03b} Babichev, E., Dokuchaev, V., and Eroshenko, Yu.
Zh. Eksp. Teor. Fiz. 127, 597 (JETP, 2005 100, 528

\bibitem{FukEbiMak92} Fukushige, T., Ebisuzaki, T., and Makino,
J. 1992, Astrophys. J., 396, L61.

\bibitem{HRKHC06} Hopkins, P.F., Robertson, B., Krause, E.,
Hernquist, L., and Cox, T.J. 2006, Astrophys. J., 652, 107.

\bibitem{HaeKau02} Haehnelt, M.G., and Kauffmann, G. 2002, Mon.
Not. Roy. Astron. Soc., 336, L61.

\bibitem{ValVal89} Valtaoja, L., and Valtonen, M.J. 1989,
Astrophys. J., 343, 47.

\bibitem{VolHaaMad02} Volonteri, M., Haardt, F., and Madau, P.
2003, Astrophys. J., 582, 559.

\bibitem{BroLoe02} Bromm, V., and Loeb, A. 2003, Astrophys. J.,
596, 34.

\bibitem{IslTaySil03} Islam, R.R., Taylor, J.E., and Silk, J.
2004, Mon. Not. Roy. Astron. Soc., 354, 427.

\bibitem{cattaneo05} Cattaneo, A., Blaizot, J., Devriendt, J., and
Guiderdoni, B. 2005, Mon. Not. Roy. Astron. Soc., 364, 407.

\bibitem{VolLodNat07} Volonteri, M., Lodato, G., and Natarajan P. 2007,
{\tt arXiv:0709.0529 [astro-ph]}.

\bibitem{ballo03} Ballo, L. et al. 2004, Astrophys. J., 600, 634.

\bibitem{komossa03} Komossa, S., et al. 2003, Astrophys. J., 582,
L15.

\bibitem{smirnova06} Smirnova, A.A., Moiseev, A.V., and Afanasiev,
V.L. 2006, Astron. Lett., 32, 520.

\bibitem{springel05} Springel, V., Di Matteo, T., and Hernquist,
L. 2005, Astrophys. J., 620, L79.

\bibitem{escala05} Escala, A., Larson, R.B., Coppi, P.S., and
Mardones, D. 2005, Astrophys. J., 630, 152.

\bibitem{merritt02} Merritt, D., and Ekers, R.D. 2002, Science,
297, 1310.

\bibitem{campan05} Campanelli, M. 2005, Class. Quant. Grav., 22,
S387.

\bibitem{merr06} Merritt, D. 2006, Astrophys. J., 648, 976.

\bibitem{berczik06} Berczik, P., Merritt, D., Spurzem, R., and
Bischof, H.-P. 2006, Astrophys. J., 642, L21.

\bibitem{masjedi06} Masjedi, M. et al. 2006, Astrophys. J., 644,
54.

\bibitem {Madau2001} Madau, P., Rees, M.J. 2001, Astrophys. J.,
551, L27.

\bibitem {All2} Fryer, C. L., Woosley, S. E., and Heger, A. 2001,
Astrophys. J., 550, 372.

\bibitem {All2b} Larson, R.B. 2000, in Star Formation from the
Small to the Large Scale, eds. F. Favata, A. Kaas and A. Wilson, ESA
SP-445, ESA, Noordwijk, p. 13.

\bibitem {All2c} Schneider, R., Ferrara, A., Ciardi, B., Ferrari,
V., and Matarrese, S. 2000, Mon. Not. Roy. Astron. Soc., 317, 385.

\bibitem {KouBulDek04} Koushiappas, S.M., Bullock, J.S., and
Dekel, A. 2004, Mon. Not. Roy. Astron. Soc., 354, 292.

\bibitem {Beg07} Begelman, M.C. 2007, To appear in First Stars
III, proc. of the conf. in Santa Fe, NM, July 16-20, 2007, eds. B.
O'Shea, A. Heger and T. Abel. AIP Conf. Proc.; {\tt
arXiv:0709.0545 [astro-ph]}.

\bibitem{SusSasTan94} Susa, H., Sasaki, M., and Tanaka T. 1994,
Prog. Theor. Phys., 92, 961.

\bibitem{Loe93} Loeb, A. 1993, Astrophys. J., 403, 542.

\bibitem{LoeRas94} Loeb, A., and Rasio, F.A. 1994, Astrophys. J.,
432, 52.

\bibitem{EisLoe95} Eisenstein, D., and Loeb, A., 1995, Astrophys.
J., 443, 11.

\bibitem{EisLoe95-2} Eisenstein, D.J., and Loeb, A. 1995, Astrophys.
J., 439, 520.

\bibitem{GurZyb90} Gurevich, A.V., and Zybin, K.P. 1990, Zh. Eksp. Teot.
Fiz., 97, 20.

\bibitem{SIZG04} Sirota, V.A., Ilyin, A.S., Zybin, K.P., and
Gurevich, A.V. 2004, {\tt arXiv:astro-ph/0403023}.

\bibitem{IZG04} Ilyin, A.S., Zybin, K.P., and Gurevich, A.V. 2004,
JETP, 98, 1.

\bibitem{ZelVas05} Zelnikov, M.I., and Vasiliev, E.A. 2005, JETP
Lett., 81, 85.

\bibitem{All1} Haehnelt, M.G., and Rees, M.J. 1993, Mon. Not. Roy.
Astron. Soc., 263, 168.

\bibitem{All1b} Loeb, A., and Rasio, F.A. 1994, Astrophys. J., 432,
52.

\bibitem{All1c} Eisenstein, D.J., and Loeb, A. 1995, Astrophys.
J., 443, 11.

\bibitem{All1d} Bromm, V., and Loeb, A. 2003, Astrophys. J., 596,
34.

\bibitem{Carr} Carr, B. 2005, {\tt arXiv:astro-ph/0511743}.

\bibitem{Khlopov80} Khlopov, M.Yu., and Polnarev, A.G. 1980,
Phys. Lett. B 97, 383.

\bibitem{Barrow81} Barrow, J.D., and Turner, M.S. 1981, Nature, 291, 469.

\bibitem{Bond82} Bond, J.R., Kolb, E.W., and Silk, L. 1982, Astrophys.
J., 255, 341.

\bibitem{Fuk86} Fukujita, M., and Rubakov, V.F. 1986, Phys. Rev.
Lett., 56, 988.

\bibitem{Dolg87} Dolgov, A.D., Illarionov, A.F., Kardashov, N.S.,
and Novikov, I.D. 1987, Zh. Eksp. Teor. Fiz., 94, 1 (Sov. Phys.
JETP, 1988, 67, 213].

\bibitem{Kof87} Kofman, L.A., and Linde, A.D. 1987, Nucl. Phys.,
B282, 555.

\bibitem{Kof88} Kofman, L.A., and Pogosyan, D.Yu. 1988, Phys.
Lett. B 214, 508.

\bibitem{Dol91} Dolgov A.D., and Kirilova, D.P. 1991, J. Moscow
Phys. Soc, 1, 217.

\bibitem{CDol92} Chizhov, M.V., and Dolgov, A.D. 1992, Nucl. Phys.
B327, 527.

\bibitem{Iok91} Yokoyama J., Suto, Y. 1991, Astrophys. J., 379,
427.

\bibitem{DolgovSilk93} Dolgov A., and Silk, L. 1993, Phys. Rev. D 47,
4244.

\bibitem{Star92} Starobinsky, A.A. 1992, Pisma Zh. Eksp. Teor.
Fiz., 55, 477 (Sov. JETP Lett., 1992, 55, 489).

\bibitem{ivan94} Ivanov, P., Naselsky,  P., and  Novikov, I. 1994,
Phys. Rev. D., 50, 7173.

\bibitem{Yoko} Yokoyama, J. 1998, Phys. Rev. D, 58, 083510.

\bibitem{Jedamzik} Jedamzik, K. 1997, Phys. Rev. D, 55, 5871.

\bibitem{Jedamzik2} Jedamzik, K. 1998, Phys. Rep., 307, 155.

\bibitem {Bubble} Crawford, M., and Schramm, D.N. 1982, Nature, 298,
538.

\bibitem {Bubble2} Hawking, S.W., Moss, I., and Stewart, J. 1982,
Phys. Rev. D, 26, 2681.

\bibitem {Bubble3} Kodama, H., Sato, K., and Sasaki, M. 1982,
Prog. Theor. Phys., 68, 1979.

\bibitem{Bubble4} La, D., and Steinhardt, P.J. 1989, Phys. Lett.
B., 220, 375.

\bibitem{Bubble5} Moss, I.G. 1994, Phys. Rev. D, 50, 676.

\bibitem{Bubble6} Khlopov, M.Yu., Konoplich, R.V., Rubin, S.G.,
and Sakharov, A.S. 1999, Grav. Cosm., 2, 1.

\bibitem{Bubble7} Konoplich, R.V., Rubin, S.G., Sakharov, A.S.,
and Khlopov, M.Yu. 1999, Phys. Atom. Nuc., 62, 1593.

\bibitem{Ru25} Dymnikova, I., Kozel, L., Khlopov, M. and Rubin, S. 2000
Grav.\& Cosm., vol.~6, pp.~311--318.

\bibitem{Ru1} Rubin, S.G., Khlopov, M.Y., and Sakharov, A.S.
2000, Grav.\& Cosm., S6, 51.

\bibitem{Ru01b} Rubin, S.G., Khlopov. M.Y., and Sakharov, A.S.
2001, JETP, 92, 921

\bibitem{Ru05} Khlopov. M.Y., Rubin, S.G., and Sakharov, A.S.,
2005, Astrop. Phys., 23, 265.

\bibitem{niem98} Niemeyer, J.C., and Jedamzik, K. 1998, Phys. Rev.
Lett., 80, 5481.

\bibitem{niem99} Niemeyer, J.C., and Jedamzik, K. 1999, Phys. Rev.
D 59, 124013.

\bibitem{niem99b} Niemeyer, J.C., and Jedamzik, K. 1999, Phys. Rev.
D 59, 124014.

\bibitem{chop} Choptuik, M.W. 1993, Phys. Rev. Lett., 70, 9.

\bibitem{yok98} Yokoyama, J. 1998, Phys. Rev. D, 58 107502.

\bibitem{DokEro01} Dokuchaev, V.I., and Eroshenko, Yu.N. 2002,
Zh. Eksp. Teot. Fiz, 121, 5 (JETP, 2002, 94, 1).

\bibitem{Dvali} Dvali, G., and Kachru, S. 2003, {\tt
arXiv:hep-th/0309095}.

\bibitem{Racetrack} Blanco-Pillado, J.J. et al. 2004, JHEP, 0411,
063.

\bibitem{LindeHyb} Linde, A. 1991, Phys. Lett. B 259, 38.

\bibitem{Dolgov97} Dolgov, A., Freese, K., Rangarajan, R., and
Srednicki, M. 1997, Phys. Rev. D, 56, 6155.

\bibitem {Star} Starobinsky, A. 1986, in Field Theory,
Quantum Gravity and Strings, ed. H.J.~de Vega and N.~Sanchez, 107.

\bibitem {Book1} Khlopov, M.Yu., and Rubin, S.G. 2004, Cosmological
Pattern of Microphysics in the Inflationary Universe, Kluwer
Academic, Dordrecht, Netherlands.

\bibitem {PBH} Rubin, S.G. 2003, in Proc. conf. I.~Ya. Pomeranchuk and
physics at the turn of the century, Moscow, p. 413, {\tt
arXiv:astro-ph/0511181}.

\bibitem {DER} Dokuchaev, V., Eroshenko, Y., and Rubin, S. 2005,
Grav. Cosm. 11, 41.

\bibitem{Zeldovich74} Zeldovich, Ya.B., Kobzarev, I.Iu., and
Okun, L.B. 1974, Zh. Eksper. Teor. Fiz., 67, 3 (Sov. Phys. ---
JETP, 1975, 40, 1).

\bibitem {TolMc30} Tolman, R.C. 1930, Phys. Rev., 35, 875.

\bibitem{kt} Kolb, E.W., and  Tkachev, I.I. 1994, Phys. Rev. D, 50,
769.

\bibitem{Carr94} Carr, B.J., Gilbert, J.H., and Lidsey J.E. 1994,
Phys. Rev. D, 50, 4853.

\bibitem{Schw03} Schwarz, D.J. 2003, Annalen Phys., 12, 220.

\bibitem{merging} Koushiappas, S.M., and Zentner, A.R. 2006,
Astrophys. J., 639, 7.

\bibitem{merging2} Koushiappas, S.M., Bullock, J.S., and Dekel, A.
2004, Mon. Not. Roy. Astron. Soc., 354, 292.

\bibitem{merging3} Volonteri, M., Haardt, F., and Madau, P. 2003,
Astrophys. J., 582, 559.

\bibitem{merging4} Bromm, V., and Loeb, A. 2003, Astrophys. J.,
596, 34.

\bibitem{Deb} Debattista, V.P., et al. 2006, Astrophys. J., 651,
L97.

\end{thebibliography}
\end{document}